\documentclass[useAMS, usenatbib]{mnras}
\usepackage{graphicx,amsmath,color,amssymb}

\usepackage[pdftitle={}]{hyperref}

\newcommand{\beq}{\begin{equation}}
\newcommand{\eeq}{\end{equation}}
\newcommand{\barr}{\begin{eqnarray}}
\newcommand{\earr}{\end{eqnarray}}

\newcommand{\Ly}{\textrm{Ly}}
\newcommand{\Lya}{Lyman-$\alpha$}
\newcommand{\Lyb}{Lyman-$\beta$}
\newcommand{\dla}{\textsc{dla}}

\newcommand{\Data}{\mathcal{D}}
\newcommand{\model}{\mathcal{M}}
\newcommand{\gp}{\textsc{gp}}

\newcommand{\realspace}{\mathbb{R}}
\newcommand{\Prob}{\textrm{Pr}}
\newcommand{\jointGaussian}{{\mathcal{N}}(\boldsymbol{\mu}, \boldsymbol{\Sigma}) }
\newcommand{\GP}{\mathcal{GP}}
\newcommand{\diag}{\textrm{diag\;}}
\newcommand{\nan}{\textrm{NaN}}
\newcommand{\normal}{\mathcal{N}}
\newcommand{\indicator}{\mathbb{I}}
\newcommand{\dd}{\textrm{d}}

\newcommand{\mdla}{\mathcal{M}_{\textrm{DLA}}}
\newcommand{\mnodla}{\mathcal{M}_{\neg\textrm{DLA}}}
\newcommand{\mkdla}{\mathcal{M}_{\textrm{DLA(k)}}}
\newcommand{\msubdla}{\mathcal{M}_{\textrm{sub}}}
\newcommand{\meanflux}{\mu(\boldsymbol{z})}
\newcommand{\pidla}{p^{i}_\textrm{DLA}}
\newcommand{\pjdla}{p^{j}_\textrm{DLA}}
\newcommand{\pdla}{p_\textrm{DLA}}
\newcommand{\midlak}{\{\mathcal{M}_{\textrm{DLA}(i)}\}_{i=1}^{k}}
\newcommand{\DLA}{\textrm{DLA}}

\newcommand{\sdss}{\textsc{sdss}}
\newcommand{\boss}{\textsc{boss}}
\newcommand{\lambdarest}{\lambda_{\textrm{rest}}}
\newcommand{\lambdaobs}{\lambda_{\textrm{obs}}}
\newcommand{\zqso}{z_{\textrm{QSO}}}
\newcommand{\qso}{\textsc{qso}}
\newcommand{\lambdavec}{\boldsymbol{\lambda}}
\newcommand{\yvec}{\boldsymbol{y}}
\newcommand{\muvec}{\boldsymbol{\mu}}
\newcommand{\fvec}{\boldsymbol{f}}
\newcommand{\Kvec}{\boldsymbol{K}}
\newcommand{\nuvec}{\boldsymbol{\nu}}
\newcommand{\Vvec}{\boldsymbol{V}}
\newcommand{\omegavec}{\boldsymbol{\omega}}
\newcommand{\Omegavec}{\boldsymbol{\Omega}}
\newcommand{\AAtext}{\textrm{\AA}}

\newcommand{\effectivetau}{\tau_{\textrm{eff,HI}}}
\newcommand{\effectivetauvec}{\boldsymbol{\tau}_{\textrm{eff,HI}}}
\newcommand{\zvec}{\boldsymbol{z}}
\newcommand{\Yvec}{\boldsymbol{Y}}
\newcommand{\Mvec}{\boldsymbol{M}}

\newcommand{\alyavec}{\boldsymbol{a}_{\textrm{F}}}
\newcommand{\avec}{\boldsymbol{a}}
\newcommand{\Alyavec}{\boldsymbol{A}_{\textrm{F}}}
\newcommand{\Avec}{\boldsymbol{A}}
\newcommand{\zmin}{z_\textrm{min}}
\newcommand{\zmax}{z_\textrm{max}}

\newcommand{\omegadla}{\Omega_{\textrm{DLA}}}
\newcommand{\lognhi}{\log_{10}{N_{\textrm{HI}}}}
\newcommand{\nhi}{N_{\textrm{HI}}}
\newcommand{\zdla}{z_{\textrm{DLA}}}
\newcommand{\kms}{\,\textrm{km\,s}^{-1}}
\newcommand{\cm}{\,\textrm{cm}}

\newcommand{\mapval}{\textsc{map}}
\newcommand{\dr}{\textsc{dr}}
\newcommand{\uvb}{\textsc{uvb}}
\newcommand{\cddf}{\textsc{cddf}}
\newcommand{\roc}{\textsc{roc}}
\newcommand{\auc}{\textsc{auc}}
\newcommand{\hi}{\textsc{hi}}
\newcommand{\matlab}{\textsc{matlab}}
\newcommand{\hpcc}{\textsc{hpcc}}
\newcommand{\ucr}{\textsc{ucr}}
\newcommand{\cnn}{\textsc{cnn}}
\newcommand{\cnr}{\textsc{cnr}}
\newcommand{\snr}{\textsc{snr}}
\newcommand{\tpr}{\textsc{tpr}}
\newcommand{\fpr}{\textsc{fpr}}
\newcommand{\ectwo}{\textsc{ec2}}
\newcommand{\nonetwo}{\textsc{n12}}
\newcommand{\pwzeronine}{\textsc{pw09}}
\newcommand{\conefive}{\textsc{c15}}
\newcommand{\boldTheta}{\boldsymbol{\Theta}}
\newcommand{\pidlabin}{\pidla( \{\mdla\} \mid \boldTheta)}
\newcommand{\pjdlabin}{\pjdla( \{\mdla\} \mid \boldTheta)}

\begin{document}

\title[Multi-DLAs with GP]{Detecting Multiple DLAs per Spectrum in SDSS DR12 with Gaussian Processes}
\author[ M.-F. Ho et al.]{Ming-Feng Ho$^1$\thanks{E-mail: mho026@ucr.edu}, Simeon Bird$^1$\thanks{E-mail: sbird@ucr.edu}, Roman Garnett$^2$.\\
$^1$University of California Riverside, Riverside, CA\\
$^2$Washington University in St. Louis, St. Louis, MO\\
}

\date{\today}

\pagerange{\pageref{firstpage}--\pageref{lastpage}} \pubyear{2019}
\pagenumbering{arabic}
\label{firstpage}

\maketitle

\begin{abstract}
We present a revised version of our automated technique using Gaussian processes ({\gp}s) to detect Damped Lyman-$\alpha$ absorbers ({\dla}s) along quasar ({\qso}) sightlines. The main improvement is to allow our Gaussian process pipeline to detect multiple {\dla}s along a single sightline.
Our {\dla} detections are regularised by an improved model for the absorption from the Lyman-$\alpha$ forest which improves performance at high redshift. We also introduce a model for unresolved sub-{\dla}s which reduces mis-classifications of absorbers without detectable damping wings. We compare our results to those of two different large-scale {\dla} catalogues and provide a catalogue of the processed results of our Gaussian process pipeline using 158\,825 Lyman-$\alpha$ spectra from {\sdss} data release 12. We present updated estimates for the statistical properties of {\dla}s, including the column density distribution function ({\cddf}), line density ($\dd N/\dd X$), and neutral hydrogen density ($\omegadla$).
\end{abstract}


\section{Introduction}
Damped Ly$\alpha$ absorbers ({\dla}s) are absorption line systems with high neutral hydrogen column densities ($N_{\mathrm{HI}} > 10^{20.3} \mathrm{cm}^{-2}$) discovered in sightlines of quasar spectroscopic observations \citep{Wolfe1986}.
The gas which gives rise to \textsc{dla}s is dense enough be self-shielded from the ultra-violet background ({\uvb}) \citep{Cen2012} yet diffuse enough to have a low star-formation rate \citep{Fumagalli2015}.
{\dla}s dominate the neutral-gas content of the Universe after reionisation \citep{Gardner1997,Noterdaeme12,Zafar2013,Crighton2015}.
Simulations tell us {\dla}s are connected with galaxies over a wide range of halo masses \citep{Haehnelt1998,Prochaska1997,Pontzen2008}, and at $z\geq 2$ are formed from the accretion of neutral hydrogen gas onto dark matter halos \citep{Bird2014,Bird2015}. 
The abundance of \textsc{dla}s at different epochs of the universe ($2 < z < 5$) thus becomes a powerful probe to understand the formation history of galaxies \citep{Gardner1997, Wolfe2005}.

Finding \textsc{dla}s historically involves a combination of template fitting and visual inspection of spectra by the eyes of trained astronomers \citep{Prochaska05,Slosar11}.
Recent spectroscopic surveys such as the Sloan Digital Sky Survey (\textsc{sdss}) \citep{York2000} have taken large amount of quasar spectra ($\sim 500\,000$ in \textsc{sdss-iv} \citep{Paris2018}).
Future surveys such as the Dark Energy Spectroscopic Instrument (\textsc{desi}\footnote{\url{http://desi.lbl.gov}}) will acquire more than 1 million quasars,
making visual inspection of the spectra impractical. Moreover, the low signal-to-noise ratios of {\sdss} data makes the task of detecting {\dla}s even harder, and induces noise related detection systematics. Since the release of the {\sdss} {\dr}14 quasar catalogue \citep{Paris2018}, visual inspection is no longer performed on all quasar targets. A fully automated and statistically consistent method thus needs to be presented for current and future surveys.

We provide a catalogue of {\dla}s using {\sdss} {\dr}12 with 158\,825 quasar sightlines.
We demonstrate that our pipeline is capable of detecting an arbitrary number of {\dla}s within each spectroscopic observation, which makes it suitable for future surveys.
Furthermore, since our pipeline resides within the framework of Bayesian probability, we have the ability to make probabilistic statements about those observations with low signal-to-noise ratios. This property allows us to make probabilistic estimations of {\dla} population statistics, even with low-quality noisy data \citep{Bird17}.

Other available searches of {\dla}s in {\sdss} include: a visual-inspection survey \citep{Slosar11}, visually guided Voigt-profile fitting \citep{Prochaska05,Prochaska2009}; and three automated methods: a template-fitting method \citep{Noterdaeme12}, an unpublished machine-learning approach using Fisher discriminant analysis \citep{Concordance2012}, and a deep-learning approach using a  convolutional neural network \citep{Parks18}.
Although these methods have had some success in creating large {\dla} catalogues,
they suffer from hard-to-control systematics due to reliance either on templates or black-box training.

We present a revised version of our previous automated method based on a Bayesian model-selection framework \citep{Garnett17}.
In our previous model \citep{Garnett17}, we built a likelihood function for the quasar spectrum, including the continuum and the non-{\dla} absorption, using Gaussian processes \citep{Rasmussen05}.
The {\sdss} {\dr}9 concordance catalogue was applied to learn the covariance of the Gaussian process model. 
In this paper, we use the effective optical depth of the Lyman-series forest to allow the mean model of the likelihood function to be adjustable to the mean flux of the quasar spectrum, which reduces the probability of falsely fitting high-column density absorbers at high redshifts. 
We also improve our knowledge of low-column density absorbers and build an alternative model for sub-{\dla}s, 
which are the {\hi} absorbers with $19.5 < \lognhi < 20$.
These modifications allow us to extend our previous pipeline to detect an arbitrary number of {\dla}s within each quasar sightline without overfitting.

Alongside the revised {\dla} detection pipeline, we present the new estimates of {\dla} statistical properties at $z > 2$.
Since the neutral hydrogen gas in {\dla}s will eventually accrete onto galactic haloes and fuel the star formation,
these population statistics can give an independent constraint on the theory of galaxy formation.
Our pipeline relies on a well-defined Bayesian framework and contains a full posterior density on the column density and redshift for a given {\dla}.
We thus can properly propagate the uncertainty in the properties of each {\dla} spectrum to population statistics of the whole sample.
Additionally, we are also able to account for low signal-to-noise ratio samples in our population statistics since the uncertainty will be reflected in the posterior probability.
We thus substantially increase the sample size in our measurements by including these noisy observations.

\section{Notation}
We will briefly recap the notation we defined in \cite{Garnett17}.
Imagine we are observing a {\qso} with a known redshift $\zqso$.
The underlying true emission function $f(\lambdarest)$ ($f\colon \mathcal{X} \rightarrow \realspace$) of the {\qso} is a mapping relation from rest-frame wavelength to flux.
We will always assume the $\zqso$ is known and rescale the observed-frame wavelength $\lambdaobs$ to the rest-frame wavelength with $\lambdarest (= \lambdaobs / (1 + \zqso))$.
We will use $\lambda$ to replace $\lambdarest$ in the rest of the text because we only work on $\lambdarest$.

The quasar spectrum observed is not the intrinsic emission function $f(\lambda)$.
Both the instrumental noise and absorption due to the intervening intergalactic medium along the line of sight will affect the observed flux.
We thus denote the observed flux as a function $y(\lambda)$.

For a real spectroscopic observation, we measure the function $y(\lambda)$ on a discrete set of samples $\lambdavec$.
We thus denote the observed flux as a vector $\yvec$, which is defined as $y_i = y(\lambda_i)$ with $i$ representing $i^\textrm{th}$ pixel.
For a given {\qso} observation, we use $\Data$ to represent a set of discrete observations $(\lambdavec, \yvec)$.

We exclude missing values of the spectroscopic observations in our calculations.
These missing values are due to pixel-masking in the spectroscopic observations (e.g., bad columns in the CCD detectors).
We will use NaN (`not a number') to represent those missing values in the text, and we will always ignore NaNs in the calculations.

\section{Bayesian Model Selection}
The classification approach used in our pipeline depends on Bayesian model selection.
Bayesian model selection allows us to compute the probability that a spectroscopic sightline $\Data$ contains an arbitrary number of {\dla}s through evaluating the probabilities of a set of models $\{\model_i\}$, where $i$ is a positive integer.
This set of $\model_i$ contains all potential models we want to classify: a model with no {\dla} and models having between one {\dla} and $k$ {\dla}s.

For each $\model_i$,
we want to compute the probability that best explains the data $\Data$ given a model $\model$.
To do this, we have to marginalize the model parameters $\theta$ and evaluate the model evidence,
\begin{equation}
   p(\Data \mid \model) =
   \int p(\Data \mid \model, \theta) p(\theta \mid \model) d\theta.
   \label{eq:marginal_params}
\end{equation}
Given a set of model evidences $p(\Data \mid \model_i)$ and model priors $\Prob (\model_i)$,
we are able to evaluate the posterior of a model given data based on Bayes's rule,
\begin{equation}
   \Prob(\model \mid \Data) =
   \frac{p(\Data \mid \model)
      \Prob(\model)}{
      \sum_i p(\Data \mid \model_i)\Prob(\model_i)
   }.
   \label{eq:model_selection}
\end{equation}
We will select the model from $\{M_i\}$ with the highest posterior.
Readers may think of this method as an application of Bayesian hypothesis testing.
Instead of only getting the likelihoods conditioned on models,
we get posterior probabilities for each model given data.

Let $k$ be the maximum number of {\dla}s we will want to detect in a quasar spectrum.
For our multi-{\dla} model selection, we will develop $k + 2$ models, which include a null model for no {\dla} detection ($\mnodla$), models for detecting exactly $k$ {\dla}s ($\mkdla$), and a model with sub-{\dla}s ($\msubdla$).
With a given spectroscopic sightline $\Data$, we will compute the posterior probability of having exactly $k$ {\dla}s in data $\Data$, $\Prob (\mkdla \mid \Data)$.

\section{Gaussian Processes}
\label{sec:gp}
In this section, we will briefly recap how we use {\it Gaussian processes} ({\gp}s) to describe the {\qso} emission function $f(\lambda)$, following \cite{Garnett17}.
The {\qso} emission function is a complicated function without a simple form derived from physically motivated parameters.
We thus use a nonparametric framework, Gaussian processes, for modelling this physically unknown function $f(\lambda)$.
A detailed introduction to {\gp}s may be found in \cite{Rasmussen05}.

\subsection{Definition and prior distribution}
\label{subsec:prior}
We wish to use a Gaussian process to model the {\qso} emission function $f(\lambda)$.
We can treat a Gaussian process as an extension of the joint Gaussian distribution $\jointGaussian$ to infinite continuous domains.
The difference is that a Gaussian process is a distribution over functions, not just a distribution over a finite number of
random variables (although since we are dealing with pixelised variables here the distinction is less important). 

A {\gp} is completely specified by its first two central moments,
a mean function $\mu(\lambda)$ and a covariance function $K(\lambda, \lambda')$:
\begin{equation}
   \begin{split}
      \mu(\lambda) &= \mathbb{E}\left[ f(\lambda) \mid \lambda \right],\\
      K(\lambda, \lambda') &= \mathbb{E}\left[ (f(\lambda) - \mu(\lambda)) (f(\lambda') - \mu(\lambda')) \mid \lambda, \lambda' \right]\\
      &= \textrm{cov} \left[ f(\lambda), f(\lambda') \mid \lambda, \lambda' \right].
   \end{split}
   \label{eq:gp_mean_cov}
\end{equation}

The mean vector describes the expected behaviour of the function, and the covariance function specifies the covariance between pairs of random variables.
We thus will write the {\gp} as,
\begin{equation}
   f(\lambda) \sim \GP(\mu(\lambda), K(\lambda,\lambda')).
   \label{eq:gp}
\end{equation}
We can write the prior probability distribution of a {\gp} as,
\begin{equation}
   p(f) = \GP(f; \mu, K).
   \label{eq:gp_prob}
\end{equation}

Real spectroscopic observations measure a discrete set of inputs $\lambdavec$ and the corresponding $f(\lambdavec)$, so we get a multivariate Gaussian distribution
\begin{equation}
   p(\boldsymbol{f}) = \mathcal{N}(f(\lambdavec); \mu(\lambdavec), K(\lambdavec, \lambdavec') ).
   \label{eq:emission_GP}
\end{equation}
Assuming the dimension of $\lambdavec$ and $\boldsymbol{f}$ is $d$, the form of the multivariate Gaussian distribution is written as
\begin{equation}
   \begin{split}
      \mathcal{N}(\fvec; \muvec, \Kvec) =& \\
      \frac{1}{\sqrt{(2\pi)^d \textrm{det}\Kvec }}
      &\exp{
         \left( -\frac{1}{2} (\fvec - \muvec)^\top \Kvec^{-1} (\fvec - \muvec) \right).
      }
   \end{split}
   \label{eq:joint_gaussian_d}
\end{equation}

\subsection{Observation model}
\label{subsec:observation_model}
We now have a Gaussian process model for a discrete set of wavelengths $\lambdavec$ and true emission fluxes $\fvec$.
To build the likelihood function for observational data $\Data = (\lambdavec, \yvec)$, we have to incorporate the observational noise.
Here we assume the observational noise is modelled by an independent Gaussian variable for each wavelength pixel,
allowing the noise realisation to differ between pixels but neglecting inter-pixel correlations.

The noise variance for a given $\lambda_i$ is written as $\nu_i = \sigma(\lambda_i)^2$.
$\sigma(\lambda_i)$ is the measurement error from a single observation on a given wavelength point $\lambda$.
With the above assumptions, we can write down the mechanism of generating observations as:
\begin{equation}
   p(\yvec \mid \lambdavec, \fvec, \nuvec)
   = \mathcal{N}(\yvec; \fvec, \Vvec),
\end{equation}
where $\Vvec = {\diag} \nuvec$, which means we put the vector $\nuvec$ on the diagonal terms of the diagonal square matrix $\Vvec$.

Given an observational model $p(\yvec \mid \lambdavec, \fvec, \nuvec)$ and a Gaussian process emission model $p(\fvec \mid \lambdavec)$, the prior distribution for observations $\yvec$ is obtained by marginalizing the latent function $\fvec$:
\begin{equation}
   \begin{split}
      p(\yvec \mid \lambdavec, \nuvec) &=
      \int p(\yvec \mid \lambdavec, \fvec, \nuvec) p(\fvec \mid \lambdavec)d\fvec\\
      &= \int \mathcal{N}(\yvec; \fvec, \Vvec) \mathcal{N}(\fvec; \muvec, \Kvec) d\fvec\\
      &=  \mathcal{N} (\yvec; \muvec, \Kvec + \Vvec),
   \end{split}
   \label{eq:marginalize_f}
\end{equation}
where the Gaussians are closed under the convolution.
Our observation model thus becomes a multivariate normal distribution described by a mean model $\mu(\lambdavec)$, covariance structure $K(\lambda, \lambda')$, and the instrumental noise $\Vvec$.
The instrumental noise is derived from {\sdss} pipeline noise, so it is different from {\qso}-to-{\qso};
however, since $\Kvec$ encodes the covariance structure of quasar emissions, $\Kvec$ should be the same for all quasars. 

As explained in \cite{Garnett17}, there is no obvious choice for a prior covariance function $\Kvec$ for modelling the quasar emission function.
Most off-the-shelf covariance functions assume some sort of translation invariance, but this is not suitable for spectroscopic observations\footnote{Detailed explanations are in \cite{Garnett17} Section 4.2.1.}.
However, we understand the quasar emission function will be independent of the presence of a low redshift {\dla}. We also assume that quasar emission functions are roughly redshift independent in the wavelength range of interest (Lyman limit to \Lya), as accretion physics should not strongly vary with cosmological evolution.
We thus build our own custom $\mu$ and $K$ for the {\gp} prior to model the quasar spectra.

\section{Learning A GP Prior from QSO Spectra}
\label{sec:gp_priors}
In this section, we will recap the prior modelling choices we made in \cite{Garnett17} and the modifications we made to reliably detect multiple {\dla}s in one spectrum.
We first build a {\gp} model for {\qso} emission in the absence of {\dla}s, the null model $\mnodla$.
Our model with {\dla}s ($\mdla$) extends this null model.
With the model priors and model evidence of all models we are considering, we compute the model posterior with Bayesian model selection.

The {\gp} prior is completely described by the first two moments, the mean and covariance functions, which we derive from data.
We must consider the mean flux of quasar emission, the absorption effect due to the {\Lya} forest, and the covariance structure within the Lyman series.

\subsection{Data}
\label{subsec:data}


Our training set to learn our {\gp} null model comprises the spectra observed by {\sdss} {\boss} {\dr}9 and labelled as containing (or not) a {\dla} by \cite{Lee2013}. \footnote{However, we use the {\dr}12 pipeline throughout.}
The {\dr}9 dataset includes 54\,468 {\qso} spectra with $\zqso > 2.15$.
We removed the following quasars from the training set:
\begin{itemize}
   \item \texttt{$z_\textrm{QSO} < 2.15$}: quasars with redshifts lower than $2.15$ have no {\Lya} in the {\sdss} band.
   \item \texttt{BAL}: quasars with broad absorption lines as flagged by the {\sdss} pipeline.
   \item spectra with less than 200 detected pixels.
   \item \texttt{ZWARNING}: spectra whose analysis had warnings as flagged by the {\sdss} redshift estimation. Extremely noisy spectra (the \texttt{TOO\_MANY\_OUTLIERS} flag) were kept. 
\end{itemize}



\subsection{Modelling Decisions}
\label{subsec:model_decisions}
Consider a set of quasar observations $\Data = (\lambdavec, \yvec)$; we always shift the observer's frame $\lambdaobs$ to rest-frame $\lambda$ so that we can set the emissions of Lyman series from different spectra to the same rest-wavelengths.
The assumption here is that the $\zqso$s of quasars are known for all the observed spectra, which is not precisely true for the spectroscopic data we have here.
Accurately estimating the redshift of quasars is beyond the scope of this paper, and is tackled elsewhere \citep{Fauber:inprep}.

The observed magnitude of a quasar varies considerably, based on its luminosity distance and the properties of the black hole. For the observation $\yvec$ to be described by a \gp, it is necessary to normalize all flux measurements by dividing by the median flux observed between $1310$ {\AA} and $1325$ {\AA}, a wavelength region which is unaffected by the {\Lya} forest.

We model the same wavelength range as in \cite{Garnett17}:
\begin{equation}
   \lambda  \in [ 911.75 \AAtext, 1215.75 \AAtext ],
   \label{eq:lambda_spacing}
\end{equation}
going from the quasar rest frame Lyman limit to the quasar rest frame {\Lya}.
The spacing between pixels is $\Delta \lambda = 0.25 \AAtext$.
Note that we prefer not to include the region past the Lyman limit. This is partly due to the relatively small amount of data in that region and partly because the non-Gaussian Lyman break associated with Lyman limit systems can confuse the model.
In particular, it occasionally tries to model a Lyman break with a wide {\dla} profile with a high column density. We shall see this is especially a problem if the quasar redshift is slightly inaccurate. The code considers the prior probability of a Lyman break at a higher redshift than the putative quasar rest frame to be zero and thus is especially prone to finding other explanations for the large absorption trough.

To model the relationship between flux measurements and the true {\qso} emission spectrum, we have to add terms corresponding to instrumental noise and weak Lyman-$\alpha$ absorption to the intrinsic correlations within the emission spectrum.
Instrumental noise was already added in Eq.~\ref{eq:marginalize_f} as a matrix $\Vvec$.

The remaining part of the modelling is to define the {\gp} covariance structure for quasars across different redshifts.
In \cite{Garnett17}, {\Lya} absorbers were modelled by a single additive noise term, $\Omegavec$, accounting for the effect of the forest as extra noise in the emission spectrum. This is not completely physical: it assumes that the \Lya~forest is just as likely to cause emission as absorption.

Here we rectify this by not only including the {\Lya} perturbation term in our Gaussian process as $\Omegavec$, but introducing a redshift dependent mean flux ($\mu(\zvec)$) with a dependence on the absorber redshift ($z(\lambdaobs)$).
We model the overall mean model with a redshift dependent absorption function and a mean emission vector: $\mu(\zvec) = a(\zvec) \circ \muvec$. 
The notation $\circ$ refers to Hadamard product, which is the element-wise product between two vectors or matrices.
The covariance matrix is decomposed into $\Alyavec (\Kvec + \Omegavec) \Alyavec$, where $\diag(\Alyavec) = a(\zvec)$ and $\Alyavec$ is a diagonal matrix.\footnote{$\Alyavec^\intercal = \Alyavec$ because it is diagonal.}
The $\Kvec$ matrix describes the covariance between different emission lines in the quasar spectrum, which we will learn from data. The $\Alyavec$ matrix is applied to $\Kvec$ because we assume that $\Kvec$ is learned before the absorption noise $a(\zvec)$ is applied.
See Sec\ref{subsec:learn_covariance} for how we learn the covariance.

Combining all modelling decisions, the model prior for an observed {\qso} emission is:
\begin{equation}
   \begin{split}
      p(\yvec \mid \lambdavec, \nuvec, &\zqso, \mnodla)
      =\\ &\mathcal{N}(\yvec; \meanflux, \Alyavec(\Kvec + \Omegavec)\Alyavec + \Vvec ).
   \end{split}
   \label{eq:model_evidence_null}
\end{equation}
The mean emission flux is now redshift- and wavelength-dependent, so the optimisation steps will differ slightly from \cite{Garnett17}. We will address the modifications in the following subsections.

\subsection{Redshift-Dependent Mean Flux Vector}
\label{subsec:redshift_dep_MF}
In this paper, instead of using a single mean vector $\muvec$
to describe all spectra, we adjust the mean model of the {\gp} to fit the mean flux of each quasar spectrum. For modeling the effect of forest absorption on the flux, we adopt an empirical power law with effective optical depth $\tau_0 (1 + z)^\beta$ for \Ly$\alpha$ forest \citep{Kim07}:
\begin{equation}
   \begin{split}
      a(z) = \exp{ ( -\tau_0 ( 1 + z )^{\beta} ) },
   \end{split}
   \label{eq:effective_optical_depth}
\end{equation}
where the absorber redshift $z$ is related to the observer's wavelength $\lambdaobs$ as:
\begin{equation}
   \begin{split}
      1 + z &= \frac{\lambdaobs}{\lambda_{\textrm{Ly}\alpha}}\\
      &= \frac{\lambdaobs}{1215.7 \AAtext}\\
      &= (1 + \zqso)\frac{\lambda}{1215.7 \AAtext},
   \end{split}
   \label{eq:z_absorbers}
\end{equation}
so the absorber redshift $z(\lambdaobs) = z(\lambda, \zqso)$ is a function of the quasar redshift and the wavelength.

In \cite{Garnett17}, we assumed the absorption from the forest would only play a role in the additive noise term ($\omegavec$) in our likelihood model $p(\yvec \mid \lambdavec, \nuvec, \omegavec, \zqso, \mnodla)$ with the form:
\begin{align}
   \omega'(\lambda, \lambdaobs) &= \omega(\lambda)s(z(\lambdaobs))^2;\\
   s(z) &= 1 - \exp{( -\tau_0 (1 + z)^\beta )} + c_0,
   \label{eq:absorption_noise_old}
\end{align}
where $z$ is the absorber redshift. The $\omega(\lambda)$ term represents the global absorption noise, and the $s(z)$ corresponds to the absorption effect contributed by the {\Lya} absorbers along the line of sight as a function of the absorber redshift $z$.

\begin{figure*}
   \includegraphics[width=2\columnwidth]{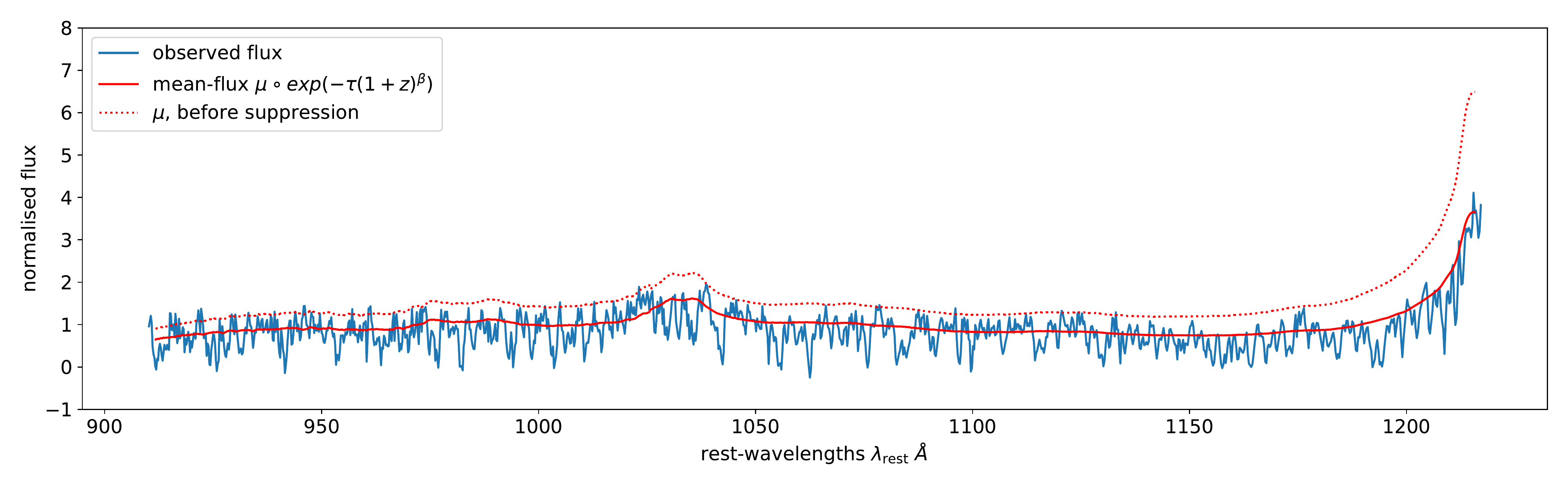}
   \caption{
   The effect of the shift to the {\gp} mean vector from the \Lya~forest effective optical depth model ($\muvec \circ \exp{( -\tau_0 (1 + \zvec)^{\beta} )}$).
   The dotted red curve shows the mean emission model before application of the forest suppression. The solid red curve is the mean model including the forest suppression.}
   \label{fig:mean_flux_15200}
\end{figure*}

Thus in our earlier model the {\Lya} forest introduces additional fluctuations
in the observed spectrum $\yvec$.
This assumption worked well for low-redshift spectra, because mean absorption due to the {\Lya} forest at low redshifts is relatively small. At high-redshifts however, the suppression of the mean flux induced by many {\Lya} absorbers is substantial, see Figure~\ref{fig:mean_flux_15200}.
In our earlier model, essentially all high-redshift {\qso} spectra were substantially more absorbed than the mean emission model $\mu$ due to absorption from the \Lya~forest. To explain this absorption, our model would fit multiple {\dla}s with large column densities.

We have improved the modelling of the \Lya~forest by allowing the mean \gp~model $\mu$ to be redshift dependent, having a mean optical depth following the measurement of \cite{Kim07}:
\begin{equation}
   \begin{split}
      \tau_{\textrm{eff}}(z) &= \tau_0 (1 + z)^{\gamma}\\
      &= 0.0023 \times \exp{(1 + z)^{3.65}},
   \end{split}
   \label{eq:kim_effective_tau}
\end{equation}
There are other measurements of $\tau_{\textrm{eff}}$ at higher precision than \cite{Kim07}, \citep[e.g.~][]{Becker:2013}. However, they are derived from {\sdss} data while \cite{Kim07} was derived from high resolution spectra. We therefore choose to use \cite{Kim07} to preserve the likelihood principle that priors should not depend on the dataset in question.

We include the effect of the whole Lyman series with a similar model, but however accounting for the different atomic coefficients of the higher order Lyman lines:
\begin{equation}
   \begin{split}
      \effectivetau(&z(\lambdaobs); \gamma, \tau_0) =\\
      \sum_{i=2}^{N} \tau_0 &\frac{\lambda_{1i} f_{1i}}{\lambda_{12} f_{12}}  (1 + z_{1i}(\lambdaobs))^{\gamma}
      \times I_{(z_{1i}(\min{(\lambdaobs)}), \zqso )}(z)
   \end{split}
   \label{eq:kim_effective_tau_beta}
\end{equation}
Here $f_{1i}$ represents the oscillator strength and $\lambda_{1i}$ corresponds to the transition wavelength from the $n=1$ to $n=i$ atomic energy level.
We model the Lyman series up to $N = 32$, with $i = 2$ being \Ly$\alpha$ and  $i=3$ \Ly$\beta$.
The absorption redshift $z_{1i}$ for the $n=1$ to $n=i$ transition is defined by:
\begin{equation}
   1 + z_{1i} = \frac{\lambdaobs}{\lambda_{1i}}\,.
\end{equation}
The optical depth at the line center is estimated by:
\begin{equation}
   \tau_0 = \sqrt{\pi} \frac{e^2}{m_e c}
   \frac{N_\ell f_{\ell u} \lambda_{\ell u}}{b},
   \label{eq:tau_line_cetner}
\end{equation}
where $\ell$ indicates the lower energy level and $u$ is the upper energy level.
For {\Lya}, we have $\lambda_{\ell u} = 1215.7$ {\AA} and $f_{\ell u} = 0.4164$; for {\Lyb}, we have $\lambda_{\ell u} = 1025.7$ and $f_{\ell u} = 0.07912$.
Given Eq.~\ref{eq:tau_line_cetner}, we have the effective optical depth for the {\Lyb} forest:
\begin{equation}
   \tau_{\beta} =
   \frac{f_{31} \lambda_{31}}{f_{21} \lambda_{21}} \tau_0 =
   \frac{0.07912 \times 1025.7}{0.4164 \times 1215.7}\times 0.0023 = 0.0004.
\end{equation}

The mean prior of the {\gp} model for each spectrum is re-written as:
\begin{equation}
   \begin{split}
      \muvec(z)
      &= \muvec \circ \exp{( -\effectivetau(\zvec; \gamma=3.65, \tau_0 = 0.0023) )}.
   \end{split}
   \label{eq:mean_flux_kim_prior}
\end{equation}
We will simply write $\effectivetau(\zvec) = \effectivetau(\zvec; \gamma=3.65, \tau_0 = 0.0023)$ in the following text for simplicity.
The new $\muvec$ is estimated via:
\begin{equation}
   \muvec = \frac{1}{N_{\neg \nan}} \sum_{y_{ij} \neq \nan} y_{ij} \cdot \exp{(+ \effectivetau(z_{ij})) }.
   \label{eq:mean_model}
\end{equation}
Eq.~\ref{eq:mean_model} rescales the mean observed fluxes back to the expected continuum before the suppression due Lyman series absorption, hopefully recovering approximately the true {\qso} emission function $\fvec$. Figure~\ref{fig:mean_flux_15200} shows the re-trained mean quasar emission model for an example quasar. The mean model, $\muvec$, is much closer to the peak emission flux above the absorbed forest.

\begin{figure*}
   \includegraphics[width=2\columnwidth]{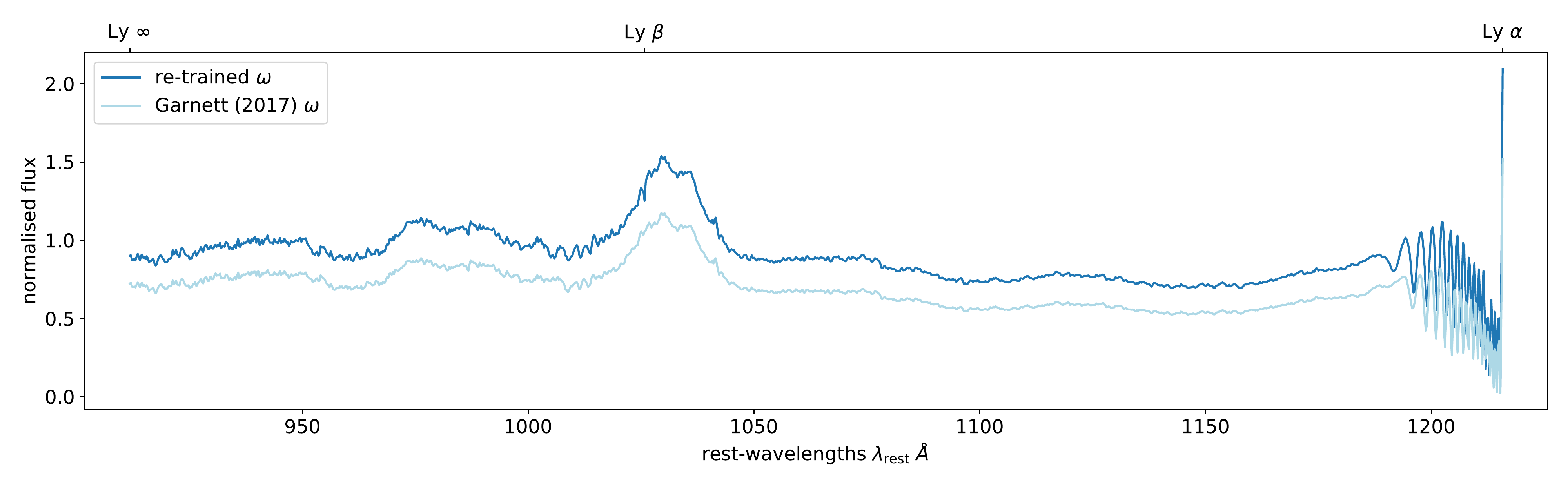}
   \caption{
   The difference between the original pixel-wise noise variance $\omegavec$ \citep{Garnett17} and the re-trained $\omegavec$  from Eq.~\ref{eq:likelihood_whole_dataset_modified}.
   The re-trained $\omegavec$ decreases because the fit no longer needs to account for the mean forest absorption.}
   \label{fig:mu_changes}
\end{figure*}

For model consistency, we account for the mean suppression from weak absorbers in our redshift-dependent noise model $\omega$ with:
\begin{align}
   \omega'(\lambda, \lambdaobs) &= \omega(\lambda)s_F(z(\lambdaobs))^2;\\
   \mathrm{where}\;s_F(z(\lambdaobs)) &= 1 - \exp{( -\effectivetau(z(\lambdaobs); \beta, \tau_0) )} + c_0\,.
   \label{eq:absorption_noise}
\end{align}
$\tau_0$, $\beta$, and $c_0$ are parameters that are learned from the data.  Figure~\ref{fig:mu_changes} shows the mean model and absorption noise variance we use, compared to the model from \cite{Garnett17}.

Note that the mean flux model introduces degeneracies between the parameters of Eq.~\ref{eq:absorption_noise}. For example, $c_0$ may be compensated by the overall amplitude of pixel-wise noise vector $\omegavec$.
For this reason, we should not ascribe strict physical interpretations to the optimal values of Eq.~\ref{eq:absorption_noise}. The optimised $\omegavec'$ is simply an empirical relation modeling the pixel-wise and redshift-dependent noise in the null model given \sdss~data.

After introducing the effective optical depth into our {\gp} mean model,
we decrease the number of large {\dla}s we detect at high redshifts and thus measure lower $\omegadla$ at high redshifts (see Section~\ref{subsec:statistical_dlas} for more details). This is because, for high redshift quasars, the mean optical depth may be close to unity. To explain this unexpected absorption, the previous code will fit multiple high-column density absorbers to the raw emission model, artificially increasing the number of {\dla}s detected. With the mean model suppressed, there is substantially less raw absorption to explain, and so this tendency is avoided.

\subsection{Learning the flux covariance}
\label{subsec:learn_covariance}
$\Kvec$ and $\Omega$ (Eq.~\ref{eq:model_evidence_null}) are optimised to maximise the likelihood of generating the data, $\Data$.
The mean flux model is not optimized, but follows the effective optical depth reported in \cite{Kim07}. Thus we remove the effect of forest absorption before we train the covariance function and train on $\Data' = \{ \lambdavec, \yvec \circ \exp{(+\effectivetau(\zvec))} - \mu(\zvec) \}$ to find the optimal parameters for $\Kvec$ and $\Omega$.

We assume the same likelihood as \cite{Garnett17} for generating the whole training data set ($\Yvec$):
\begin{equation}
   \begin{split}
      p(\Yvec \mid \lambdavec, \Vvec, \Mvec, \omegavec, &\boldsymbol{z}_{\textrm{QSO}}, \mnodla)\\
      &= \prod_{i=1}^{N_{\textrm{spec}}}
      \mathcal{N} (\yvec_i; \muvec, \Kvec + \Omegavec + \Vvec_i ),
   \end{split}
   \label{eq:likelihood_whole_dataset}
\end{equation}
where $\Yvec$ means the matrix containing all the observed flux in the training data, and the product on the right hand side says we are combining all likelihoods from each single spectrum.
The noise matrix $\Omegavec = \diag{\omegavec'}$ is the diagonal matrix which represents the {\Lya} forest absorption from Eq.~\ref{eq:absorption_noise}.

$\Mvec$ is a low-rank decomposition of the covariance matrix $\Kvec$ we want to learn:
\begin{equation}
   \Kvec = \Mvec \Mvec^{\top},
   \label{eq:low_rank_decomposition}
\end{equation}
where $\Mvec$ is an ($N_{\textrm{pixels}} \times k$) matrix.
Without this low-rank decomposition, we would need to learn $N_\mathrm{pixel}^2 = 1\,217 \times 1\,217$ free parameters.
With Eq.~\ref{eq:low_rank_decomposition}, we can limit the number of free parameters to be $N_{\textrm{pixels}} \times k$,
where $k \ll N_{\textrm{pixels}}$; also, it guarantees the covariance matrix $\Kvec$ to be positive semi-definite.
Each column of the $\Mvec$ can be treated as an eigenspectrum of the training data,
where we set the number of eigenspectra to be $k = 20$.
We will optimise the $\Mvec$ matrix and the absorption noise in Eq.~\ref{eq:absorption_noise} simultaneously.

A modification performed in this work is to, instead of directly training on the observed flux, optimise the covariance matrix and noise model on the flux with Lyman-$\alpha$ forest absorption removed (de-forest flux):
\begin{equation}
   \begin{split}
      \yvec  &:= \yvec \circ \exp{ (+\effectivetau(\zvec)) };\\
      Y_{ij} &:= Y_{ij} \exp{ (+\effectivetau(\zvec)) }_{ij}.
   \end{split}
   \label{eq:de_absorbed_flux}
\end{equation}
We may write this change into the likelihood:
\begin{equation}
   \begin{split}
      p(\Yvec &\circ \exp{(+\effectivetau(\zvec))} \mid \lambdavec, \Vvec, \Mvec, \omegavec, \boldsymbol{z}_{\textrm{QSO}}, \mnodla)\\
      &= \prod_{i=1}^{N_{\textrm{spec}}}
      \mathcal{N} (\yvec_i \circ \exp{(+{\effectivetau(\zvec_i)})}; \muvec, \Kvec + \Omegavec + \Vvec_i ),
   \end{split}
   \label{eq:likelihood_whole_dataset_modified}
\end{equation}
where $\muvec$ is the mean model from Eq.~\ref{eq:mean_model}.
The rest of our optimisation procedure follows the unconstrained optimisation of \cite{Garnett17}.

We use de-forest fluxes for training as we want our covariance matrix to learn the covariance in the true emission function.
The emission function (like our kernel $\Kvec$) is independent of quasar emission redshift, whereas the absorption noise is not.
We only implement the mean forest absorption of \cite{Kim07}, so we need an extra term to compensate for the variance of the forest around this mean.
We thus still train the redshift- and wavelength-dependent absorption noise from data. The optimal values we learned for Eq.~\ref{eq:absorption_noise} are:
\begin{align}
   c_0 = 0.3050; \tau_0 = 1.6400 \times 10^{-4}; \beta = 5.2714.
   \label{eq:learned_absorption_params}
\end{align}
As we might expect, the optimal $\tau_0$ value is smaller than the $\tau_0 = 0.01178$ learned in \cite{Garnett17}, which implies the effect of the forest is almost removed by applying the Lyman- series forest to the mean model.

\begin{figure}
   \centering
   \includegraphics[width=1\columnwidth]{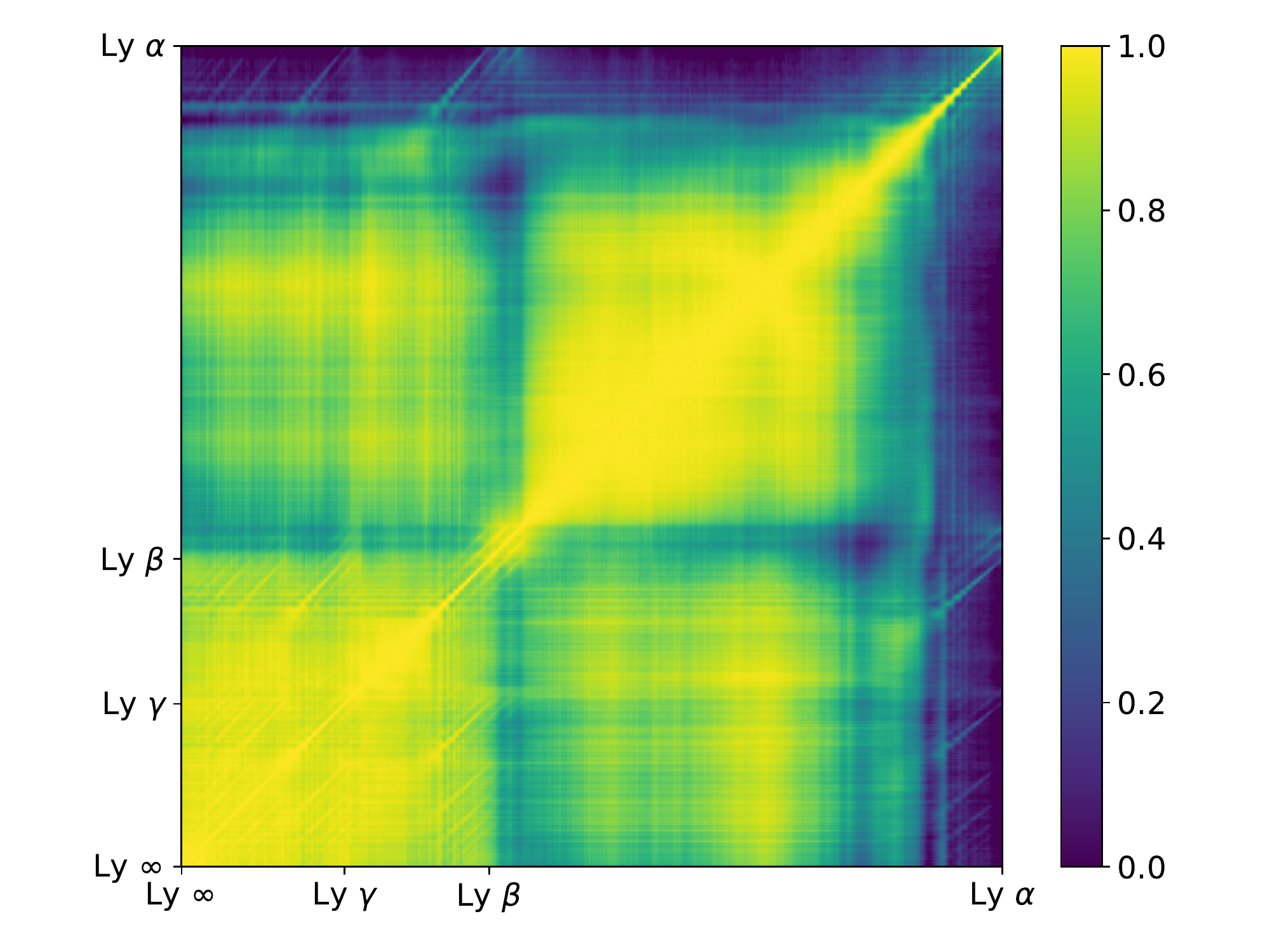}
   \caption{The trained covariance matrix $\Mvec$, which is almost the same as the covariance from \protect\cite{Garnett17}.
   Note that we normalize the diagonal elements to be unity, so this is more like a correlation matrix than a covariance matrix.
   The values in the matrix are ranging from 0 to 1, representing the correlation between $\lambda$ and $\lambda'$ in the {\qso} emission.}
\end{figure}

\subsection{Model evidence}
\label{subsec:model_evidence_null}
Consider a given {\qso} observation $\Data = (\lambdavec, \yvec)$ with known observational noise $\nu(\lambdavec)$ and known {\qso} redshift $\zqso$.
The model evidence for $\mnodla$ can be estimated using
\begin{equation}
   p(\Data \mid \mnodla, \nu, \zqso) \propto
   p(\yvec \mid \lambdavec, \nuvec, \zqso, \mnodla),
   \label{eq:model_evidence_null_model}
\end{equation}
which is equivalent to evaluating a multivariate Gaussian
\begin{equation}
   \begin{split}
      p(\yvec &\mid \lambdavec, \nuvec, \zqso, \mnodla) \\
      &= \normal(\yvec; \mu \circ \exp{ (-\effectivetauvec) }, \Alyavec(\Kvec + \Omegavec)\Alyavec + \Vvec )\,.
   \end{split}
   \label{eq:model_evidence_null_gaussian}
\end{equation}
Here $\exp{(-\effectivetauvec)} = \diag{\Alyavec}$ describes the absorption due to the forest and modifies the mean vector $\muvec$, the covariance matrix $\Kvec$ and the noise matrix $\Omegavec$ to account for the \Lya~forest effective optical depth.

\section{A GP Model for QSO Sightlines with Multiple DLAs}
\label{sec:model_multi_dlas}
In Section~\ref{sec:gp_priors}, we learned a {\gp} prior for {\qso} spectroscopic measurements without any {\dla}s for our null model $\mnodla$.
Here we extend the null model $\mnodla$ to a model with $k$ intervening {\dla}s, $\mkdla$.

Our complete {\dla} model, $\mdla$, will be the union of the models with $i$ {\dla}s:  $\mdla = \{ {\mdla}_{(i)} \}_{i=1}^{k}$.
We consider only until $k=4$, as {\dla}s are rare events and our sample only contains one spectrum with $4$ {\dla}s.

\subsection{Absorption function}
\label{subsec:voigt_profile}
Before we model a quasar spectrum with intervening {\dla}s, we need to have an absorption profile model for a {\dla}.
Damped Lyman alpha absorbers, or {\dla}s, are neutral hydrogen ({\hi}) absorption systems with saturated lines and damping wings in the spectroscopic measurements.
Having saturated lines means the column density of the absorbers on the line of sight is high enough to absorb essentially all photons.
The damping wings are due to natural broadening in the line.

The optical depth from each Lyman series transition is
\begin{equation}
   \tau(\lambda; \zdla, \nhi)
   = \nhi \frac{\pi e^2 f_{1u} \lambda_{1u}}{m_e c}
   \phi(v, b, \gamma),
\end{equation}
where $e$ is the elementary charge, $\lambda_{1u}$ is the transition wavelength from the $n=1$ to $n = u$ energy level ($\lambda_{12} = 1215.6701$ {\AA} for {\Lya}) and $f_{1u}$ is the oscillator strength of the transition.
The line profile $\phi$ is a Voigt profile:
\begin{equation}
   \begin{split}
      \phi(v, b, \gamma)
      &=\\
      \int \frac{d v}{\sqrt{2\pi}\sigma_v} &\exp{(-v^2 / 2\sigma_v^2)}
      \frac{4 \gamma_{\ell u}}{16 \pi^2 [ \nu - (1 - v/c)\nu_{\ell u} ]^2 + \gamma_{\ell u}^2},
   \end{split}
\end{equation}
which is a convolution between a Lorenztian line profile and a Gaussian line profile.
The $\sigma_v$ is the one-dimensional velocity dispersion, $\gamma_{\ell u}$ is a parameter for Lorenztian profile, $\nu$ is the frequency, and $u$ represents the upper energy level and $\ell$ represents the lower energy level.

Both profiles are parameterised by the relative velocity $v$, which means both profiles are distributions in the 1-dimensional velocity space:
\begin{equation}
   v = c \left( \frac{\lambda}{\lambda_{1u}} \frac{1}{(1 + \zdla)} -1 \right).
\end{equation}

The standard deviation of the Gaussian line profile is related to the broadening parameter $b = \sqrt{2} \sigma_v$, and if we assume the broadening is entirely due to thermal motion:
\begin{equation}
   b = \sqrt{ \frac{2 k T}{m_p} }.
\end{equation}
Introducing the damping constant $\Gamma  = 6.265 \times 10^8 \textrm{s}^{-1}$ for {\Lya}, we have the parameter $\gamma_{\ell u}$ to describe the width of the Lorenztian profile
\begin{equation}
   \gamma_{\ell u} = \frac{\Gamma \lambda_{\ell u}}{4\pi}.
\end{equation}

Our default {\dla} profile includes ${\Ly}\alpha$, ${\Ly}\beta$, and ${\Ly}\gamma$ absorptions. We fix the broadening parameter $b$ by setting $T = 10^4 \,\textrm{K}$, which increases the width of the {\dla} profile by $13 \kms$, small compared to the effect of the Lorenztian wings.
Thus, for a given {\qso} and a true emission function $f(\lambda)$, the function for the observed flux $y(\lambda)$ is
\begin{equation}
   y(\lambda)
   = f(\lambda) \exp{( -\tau(\lambda; \zdla, \nhi) )} \exp{(-\effectivetau(\lambdaobs))}
   + \epsilon,
\end{equation}
where $\epsilon$ is additive Gaussian noise including measurement noise and absorption noise.

Suppose we have a {\dla} at redshift $\zdla$ with column density $\nhi$.
We can model the spectrum with an intervening {\dla} by calculating the {\dla} absorption function:
\begin{equation}
   \avec = \exp{(-\tau(\lambdavec; \zdla, \nhi))}.
\end{equation}
We apply the absorption function to the {\gp} prior of $\yvec$ with
\begin{equation}
   \begin{split}
      p(\yvec &\mid \lambdavec, \nuvec, \zqso, \zdla, \nhi, \mdla ) \\
      &= \normal (\yvec ; \avec \circ ( \alyavec \circ \muvec), \Avec (\Alyavec (\Kvec + \Omegavec) \Alyavec) \Avec + \Vvec),
   \end{split}
   \label{eq:likelihood_function_model_DLA}
\end{equation}
where $\Avec = \diag \avec$.

For a model with $k$ {\dla}s with $k \in \mathbb{N}$, we simply take the element-wise product of $k$ absorption functions:
\begin{equation}
   \begin{split}
      \avec_{(k)} = \prod_{i = 1}^{k} a(\lambdavec; {\zdla}_i, {\nhi}_i); \\
      \diag \Avec_{(k)} = \avec_{(k)}.
   \end{split}
\end{equation}
The prior for $\mkdla$ would therefore be:
\begin{equation}
   \begin{split}
      p(\yvec \mid \lambdavec, \nuvec, \zqso, \{{\zdla}_i\}_{i=1}^{k}, \{{\nhi}_i\}_{i=1}^{k}, \mkdla ) = \\
      \normal (\yvec ; \avec_{(k)} \circ ( \alyavec \circ \muvec), \Avec_{(k)} (\Alyavec (\Kvec + \Omegavec) \Alyavec) \Avec_{(k)} + \Vvec).
   \end{split}
   \label{eq:dla_model_k}
\end{equation}

Here we briefly review our notations in Eq\ref{eq:dla_model_k}:
$\avec_{(k)}$, which is parameterised by $(\{{\zdla}_i\}_{i=1}^{k}, \{{\nhi}_i\}_{i=1}^{k})$, 
represents the absorption function with $k$ {\dla}s in one spectrum.
Note that each {\dla} is parameterised by a pair of $(\zdla, \nhi)$.
$\alyavec$ corresponds to the absorption function from the Lyman series absorptions,
which is derived from \cite{Kim07} in the form of Eq.~\ref{eq:mean_flux_kim_prior}.
The covariance matrix $\Kvec$ and the absorption model $\Omegavec$ are both learned from data,
as described in Section~\ref{subsec:learn_covariance}.
$\Vvec$ is the noise variance matrix given by the {\sdss} pipeline,
so each sightline would have different $\Vvec$.

\subsection{Model Evidence: DLA(1)}

The model evidence of our {\dla} model is given by the integral:
\begin{equation}
   \begin{split}
      p(\Data &\mid \mdla{_{(1)}}, \zqso) \propto\\
      \int p(&\yvec \mid \lambdavec, \nuvec, \theta, \zqso \mdla{_{(1)}})
      p(\theta \mid \zqso, \mdla{_{(1)}})
      \dd\theta,
   \end{split}
   \label{eq:model_evidence}
\end{equation}
where we integrated out the parameters, $\theta = (\zdla, \lognhi)$, with a given parameter prior $p(\theta \mid \zqso, \mdla{_{(1)}})$.

However, Eq.~\ref{eq:model_evidence} is intractable, so we approximate it with a quasi-Monte Carlo method (\textsc{qmc}). \textsc{qmc} selects $N = 10\,000$ samples with an approximately uniform spatial distribution from a Halton sequence to calculate the model likelihood, approximating the model evidence by the sample mean:

\begin{equation}
   \begin{split}
      p(D &\mid \mdla{_{(1)}}, \zqso) \simeq\\
      &\frac{1}{N} \sum_{i=1}^{N} p(\Data \mid \theta_i, \zqso, \mdla{_{(1)}}).
   \end{split}
   \label{eq:model_evidence_qmc}
\end{equation}

\subsection{Model evidence: Occam's Razor Effect for DLA(k)}
\label{sec:occams_dlak}
For higher order {\dla} models, we have to integrate out not only the nuisance parameters of the first {\dla} model $\mdla{_{(1)}}$, ($\theta_1$) but also the parameters from $\mdla{_{(2)}}$ to $\mkdla$,
\begin{equation}
   \begin{split}
      p(\Data \mid \mkdla, \zqso) &\propto\\
      \int p(\Data \mid \mkdla, &\{\theta_i\}_{i=1}^{k}) \times\\
      p(\{\theta_i\}_{i=1}^{k} &\mid \mkdla, \Data, \zqso) \dd\{\theta_i\}_{i=1}^{k},
   \end{split}
   \label{eq:integral_dla_k}
\end{equation}
which means we are marginalising $\{\theta_i\}_{i=1}^{k}$ in a parameter space with $2 \times k$ dimensions.
The parameter prior of multi-{\dla}s is a multiplication between a non-informative prior $p(\theta_i \mid \mdla{_{(1)}}, \zqso)$
and the posterior of the $(k-1)$ multi-{\dla} model,

\begin{equation}
   \begin{split}
      p(\{\theta_i\}_{i=1}^{k} \mid \mkdla, \Data, \zqso) =& \\
      p(\{\theta_i\}_{i=1}^{k-1} \mid \mdla{_{(k-1)}}, \Data, \zqso)&
      p(\theta_k \mid \zqso, \mdla{_{(1)}}).
   \end{split}
   \label{eq:parameter_prior_dla_k}
\end{equation}

We can approximate this integral using the same \textsc{qmc} method.
For example, if we want to sample the model evidence for $\mdla{_{(2)}}$,
we would need $N=10\,000$ samples for each parameter dimension $\{\theta_i\}_{i=1}^{2}$,
which results in sampling from two independent Halton sequences with $10^8$ samples in total.
If we want to sample up to $\mdla{_{(k)}}$ with $N$ samples for each $\{\theta_i\}$ from $i = 1, ..., k$,
we would need to have:
\begin{equation}
   \begin{split}
      p(&\Data \mid \mkdla, \zqso) \\
      &\simeq
      \frac{1}{N} \sum_{j^{(1)}=1}^{N}
      \frac{1}{N} \sum_{j^{(2)}=1}^{N}
      \frac{1}{N} \sum_{j^{(3)}=1}^{N} \dots
      \frac{1}{N} \sum_{j^{(k)}=1}^{N}\\
      &p(\Data \mid \mkdla, \{\theta_{1j^{(1)}}, \theta_{2j^{(2)}}, \theta_{3j^{(3)}}, \dots, \theta_{kj^{(k)}}\}, \zqso),
   \end{split}
   \label{eq:qmc_dla_k}
\end{equation}
where $\{ j^{(1)},j^{(2)},j^{(3)}, \dots ,j^{(k)} \}$ indicate the indices of \textsc{qmc} samples.
The above Eq.~\ref{eq:qmc_dla_k} is thus in principle evaluated with $N^k$ samples.

In practice, we only sample $N=10\,000$ points from $p(\{\theta_i\}_{i=1}^{k} \mid \mkdla, \Data, \zqso)$ instead of sampling $N^k$ points, as a uniform sampling of the first {\dla} model may be reweighted to cover parameter space for the higher order models. A $N^{k-1}$ factor of normalisation is thus left behind in the summation,
\begin{equation}
   \begin{split}
      p(\Data \mid & \mkdla, \zqso)  \\
      &\simeq \frac{1}{N^k}
      \sum_{j=1}^{N} p(\Data \mid \mkdla, \{\theta_{ij}\}_{i=1}^{k}, \zqso)\\
      &\simeq \frac{1}{N^{k-1}}
      \left( \frac{1}{N} \sum_{j=1}^{N} p(\Data \mid \mkdla, \{\theta_{ij}\}_{i=1}^{k}, \zqso) \right) \\
      &\simeq \frac{1}{N^{k-1}}
      \textrm{mean}_j\left( p(\Data \mid \mkdla, \{\theta_{ij}\}_{i=1}^{k}, \zqso) \right).
   \end{split}
   \label{eq:qmc_occams_razor}
\end{equation}
The additional $\frac{1}{N^{k-1}}$ factor penalises models with more parameters than needed, and can be viewed as an implementation of Occam's razor.
This Occam's razor effect is caused by the fact that all probability distributions have to be normalised to unity.
A model with more parameters, which means having a wider distribution in the likelihood space, results in a bigger normalisation factor.

The motivation for us to draw $N$ samples from the multi-{\dla} likelihood function $p(\{\theta_i\}_{i=1}^{k} \mid \mkdla, \Data, \zqso)$ is that we believe the prior density we took from the posterior density of $\mdla{_{(k-1)}}$ is representative enough even without $N^{k}$ samples.
For example, if we have two peaks in our likelihood density $p( \Data \mid \mdla{_{(1)}}, \theta_1, \zqso )$, we expect the sampling for $\theta_2$ in $p(\Data \mid \mdla{_{(2)}}, \{\theta_1, \theta_2\}, \zqso)$ would concentrate on sampling the density of the first highest peak in $p( \Data \mid \mdla{_{(1)}}, \theta_1, \zqso )$ density.
Similarly, while we are sampling for $\mdla{_{(3)}}$, we expect $\theta_3$ and $\theta_2$ would cover the first and the second-highest peaks.

To avoid multi-{\dla}s overlapping with each other,
we inject a dependence between any pair of $\zdla$ parameters.
Specifically, if any pair of $\zdla$s have a relative velocity smaller than $3\,000 \kms$, then we set the likelihood of this sample to NaN.

\subsection{Additional penalty for DLAs and sub-DLAs}
\label{sec:additional_occams_razor}
In Section~\ref{sec:occams_dlak}, we apply a penalty, Occam's razor, to regularise {\dla} models using more parameters than needed.
This effect is due to the normalization (to unity) of the evidence.

In a similiar fashion, and for a similar reason, we apply an additional regularisation factor between the non-{\dla} and {\dla} models (including sub-{\dla}s).
This additional factor ensures that when both models are a poor fit to a particular observational spectrum, the code prefers the non-{\dla} model, rather than preferring the model with more parameters and thus greater fitting freedom.
We directly inject this Occam's razor factor in the model selection:
\begin{equation}
   \begin{split}
      \Prob(&\mdla \mid \Data) =\\
      &\frac{\Prob(\mdla)p(\Data \mid \mdla)\frac{1}{N}}{
         \left(\begin{matrix}
            \Prob(\mdla)p(\Data \mid \mdla)\\
            +\Prob(\msubdla)p(\Data \mid \msubdla)\\
         \end{matrix}\right)\frac{1}{N}
         +\Prob(\mnodla \mid \Data)
      },
   \end{split}
\end{equation}
where $N = 10^4$ is the number of samples we used to approximate the parameterised likelihood functions. We evaluated the impact of this regularization factor on the area under the curve (\auc) in the receiver-operating characteristics (\roc) plot.\footnote{See Section~\ref{subsec:roc_analysis} for how we compute our {\roc} plot.} For $N = 10^4$, the {\auc} changed from $0.949$ to $0.960$. We considered other penalty values and found that the {\auc} increased up to $N=10^4$ and then plateaued.

In addition, we found by examining specific examples that this penalty regularized a relatively common incorrect {\dla} detection: finding objects in short, very noisy low redshift ($z \sim 2.2$) spectra. In these spectra our earlier model would prefer the {\dla} model purely because of its large parameter freedom. In particular a high column density {\dla}, large enough that the damping wings exceed the width of the spectrum, would be preferred. Such a fit exploits a degeneracy in the model between the mean observed flux and the {\dla} column density when the spectrum is shorter than the putative {\dla}. The Occam's razor penalty avoids these spurious fits by penalising the extra parametric freedom in the {\dla} model.

\subsection{Parameter prior}
Here we briefly recap the priors on model parameters chosen in \cite{Garnett17}.
Suppose we want to make an inference for the column density and redshift of an absorber $\theta = (\nhi, \zdla)$ from a given spectroscopic observation,
the joint density for the parameter prior would be
\begin{equation}
   p(\theta \mid \zqso, \mdla{_{(1)}}) =
   p(\nhi, \zdla \mid \zqso, \mdla{_{(1)}}).
\end{equation}
Suppose the absorber redshift and the column density are conditionally independent and the column density is independent of the quasar redshift $\zqso$:
\begin{equation}
   \begin{split}
      p(\theta \mid \zqso, &\mdla{_{(1)}}) =\\
      &p(\zdla \mid \zqso, \mdla{_{(1)}})
      p(\nhi \mid \mdla{_{(1)}})
   \end{split}
\end{equation}

We set a bounded uniform prior density for the absorber redshift $\zdla$:
\begin{equation}
   p(\zdla \mid \zqso, \mdla{_{(1)}})
   = \mathcal{U}[\zmin, \zmax],
\end{equation}
where we define the finite prior range to be
\begin{equation}
   \zmin = \textrm{max}\left\{
   \begin{matrix}
      &\frac{\lambda_{\Ly \infty}}{\lambda_{\Ly \alpha}}
      (1 + \zqso) - 1 + 3\,000\,\kms / c \\
      &\frac{\min \lambdaobs}{\lambda_{\Ly \alpha}} - 1
   \end{matrix}
   \right.
\end{equation}
\begin{equation}
   \zmax = \zqso - 3\,000 \, \kms / c;
\end{equation}
which means we have a prior belief that the center of the absorber is within the observed wavelengths.
The range of observed wavelengths is either from $\Ly \infty$ to $\Ly \alpha$ of the quasar rest-frame ($\lambdarest \in$ [911.75 \AA, 1216.75 \AA])
or from the minimum observed wavelength to $\Ly \alpha$.
We also apply a conservative cutoff of 3\,000 $\kms$ near to $\Ly \infty$ and $\Ly \alpha$.
The $-3\,000 \, \kms$ cutoff for $\zmax$ helps to avoid proximity ionisation effects due to the quasar radiation field.
Furthermore, the $+3\,000 \, \kms$ cutoff for $\zmin$ avoids a potentially incorrect measurement for $\zqso$.
An underestimated $\zqso$ can produce a Lyman-limit trough within the region of the quasar expected to contain only Lyman-series absorption, and the code can incorrectly interpret this as a {\dla}.

For the column density prior, we follow \cite{Garnett17}.
We first estimate the density of {\dla}s column density $p(\nhi \mid \mdla)$ using the {\boss} {\dr}9 {\Lya} forest sample.
We choose to put our prior on the base-10 logarithm of the column density $\log_{10}\nhi$ due to the large dynamic range of {\dla} column densities in {\sdss} {\dr}9 samples.

We thus estimate the density of logarithm column densities $p(\log_{10} \nhi \mid \mdla{_{(1)}})$ using univariate Gaussian kernels on the reported $\log_{10} \nhi$ values in {\dr}9 samples.
Column densities from {\dla}s in {\dr}9 with $N_\textrm{DLA} = 5\,854$ are used to non-parametrically estimate the logarithm $\nhi$ prior density, with:
\begin{equation}
   \begin{split}
      p_\textrm{KDE}(\log_{10} \nhi &\mid \mdla{_{(1)}}) \\
      &= \frac{1}{N_\textrm{DLA}}
      \sum_{i=1}^{N_\textrm{DLA}}
      \normal(\log_{10} \nhi; l_i, \sigma^2),
   \end{split}
\end{equation}
where $l_i$ is the logarithm column density $\log_{10} \nhi$ of the $i^\textrm{th}$ sample.
The bandwidth $\sigma^2$ is selected to be the optimal value for a normal distribution, which is the default setting for {\matlab}.

We further simplify the non-parametric estimate into a parametric form with:
\begin{equation}
   \begin{split}
      p_\textrm{KDE}(&\log_{10} \nhi = N \mid \mdla{_{(1)}}) \simeq \\
      &q(\log_{10} \nhi = N) \propto \exp{(a N^2 + b N + c)};
   \end{split}
\end{equation}
where the parameters $(a, b, c)$ for the quadratic function are fitted via standard least-squared fitting to the non-parametric estimate of density $p_\textrm{KDE}(\log_{10} \nhi\mid \mdla{_{(1)}})$ with the range $\log_{10} \nhi \in [20, 22]$.
The optimal values for the quadratic terms were:
\begin{equation}
   a = -1.2695; b = 50.863; c = -509.33;
\end{equation}
Note that we have the same values as in \cite{Garnett17}.

Finally, we choose to be conservative about the data-driven column density prior.
We thus take a mixture of a non-informative log-normal prior with the data-driven prior to make a non-restrictive prior on a large dynamical range:
\begin{equation}
   \begin{split}
      p(\log_{10} &\nhi \mid \mdla{_{(1)}}) \\
      &= \alpha q(\log_{10} \nhi = N) + (1 - \alpha) \mathcal{U}[20, 23].
   \end{split}
\end{equation}
Here we choose the mixture coefficient $\alpha = 0.97$, which favours the data-driven prior. We still include a small component of a non-informative prior so that we are able to detect {\dla}s with a larger column density than in the training set, if any are present in the larger {\dr}12 sample.
Note that $\alpha = 0.97$ is 7\% higher than the coefficient chosen in \cite{Garnett17}, which was $\alpha = 0.90$. Our previous prior slightly over-estimated the number of very large {\dla}s.


\subsection{Sub-DLA parameter prior}
As reported in \cite{Bird17}, the column density distribution function ({\cddf}) exhibited an edge feature: an over-detection of {\dla}s at low column densities ($\sim 10^{20} \cm^{-2}$).
This did not affect the statistical properties of {\dla}s as we restrict column density to $\nhi \geqslant 10^{20.3} \cm^{-2}$ for both line densities ($\dd N/\dd X$) and total column densities ($\omegadla$).
However, to make our method more robust, here we describe a complementary method to avoid over-estimating the number of low column density absorbers.

The excess of {\dla}s at $\sim 10^{20}\cm^{-2}$ is due to our model excluding lower column density absorbers such as sub-{\dla}s.
Since we limited our column density prior of {\dla}s to be larger than $10^{20} \cm^{-2}$, the code cannot correctly classify a sub-{\dla}.
Instead it correctly notes that a sub-{\dla} spectrum is more likely to be a {\dla} with a minimal column density than an unabsorbed spectrum.

To resolve our ignorance, we introduce an alternative model $\msubdla$ to account the model posterior of those low column density absorbers in our Bayesian model selection.
The likelihood function we used for sub-{\dla}s is identical to the one we built for {\dla} model $\mdla{_{(1)}}$ in Eq.~\ref{eq:likelihood_function_model_DLA}
but has a different parameter prior on the column densities $p(\log_{10} \nhi \mid \msubdla )$.
We restricted our prior belief of sub-{\dla}s to be within the range $\lognhi \in [19.5, 20]$,
and, as we do not have a catalogue of sub-{\dla}s for learning the prior density, we put a uniform prior on $\lognhi$:
\begin{equation}
   p(\lognhi \mid \msubdla)
   = \mathcal{U}[19.5, 20].
\end{equation}
We place a lower cutoff at $\lognhi = 19.5$ because the relatively noisy {\sdss} data offers limited evidence for absorbers with column densities lower than this limit.

\section{Model Priors}
Bayesian model selection allows us to combine prior information with evidence from the data-driven model to obtain a posterior belief about the detection of {\dla}s $p(\mdla \mid \Data)$ using Bayes' rule.
For a given spectroscopic observation $\Data$,
we already have the ability to compute the model evidence for a {\dla} ($p(\Data \mid \mdla)$) and no {\dla} ($p(\Data \mid \mnodla)$). However, to compute the model posteriors, we need to specify our prior beliefs in these models. Here we approximate our prior belief $\Prob(\mdla)$ using the {\sdss} {\dr}9 {\dla} catalogue.

Consider a {\qso} observation $\Data = (\lambdavec, \yvec)$ at $\zqso$. We want to find our prior belief that $\Data$ contains a {\dla}.
We count the fraction of {\qso} sightlines in the training set containing {\dla}s with redshift less than $\zqso + z'$, where $z' = 30\,000 \kms / c$ is a small constant. If $N$ is the number of {\qso} sightlines with redshift less than $\zqso + z'$, and $M$ is the number of sightlines in this set containing {\dla}s in the quasar rest-frame wavelengths range we search, then our empirical prior for $\mdla$ is:
\begin{equation}
   \Prob(\mdla \mid \zqso) = \frac{M}{N}.
\end{equation}

We can break down our {\dla} prior $\Prob(\mdla \mid \zqso)$ for multiple {\dla}s in a {\qso} sightline $\Prob(\mkdla \mid \zqso)$ via:
\begin{equation}
   \Prob(\mkdla \mid \zqso) \simeq \left( \frac{M}{N} \right)^k
   - \left( \frac{M}{N} \right)^{k + 1}.
\end{equation}
For example, $\frac{M}{N}$ represents our prior belief of having at least one {\dla} in the sightline, and $(\frac{M}{N})^2$ represents having at least two {\dla}s. $\frac{M}{N} - (\frac{M}{N})^2$ is thus our prior belief of having exactly one {\dla} at the sightline.

\subsection{Sub-DLA model prior}
\label{subsec:sub_dla_model_prior}

The column density distribution function ({\cddf}) of \cite{Bird17} exhibited an edge effect at $\lognhi \sim 20$ due to a lack of sampling at lower column densities.
We thus construct an alternative model for lower column density absorbers (sub-{\dla}s, {\dla}s' lower column density cousins) to regularise {\dla} detections.
We use the same {\gp} likelihood function as the {\dla} model $\mdla$ to compute our sub-{\dla} model evidence $p(\Data \mid \msubdla)$ but with a different column density prior $p(\lognhi \mid \msubdla)$.

There is no sub-{\dla} catalogue available for us to estimate the empirical prior directly.
We, therefore, approximate our sub-{\dla} model prior by rescaling our {\dla} model prior:
\begin{equation}
   \Prob(\msubdla \mid \zqso)
   \propto \Prob(\mdla \mid \zqso),
\end{equation}
and we require our prior beliefs to sum to unity:
\begin{equation}
   \begin{split}
      \Prob(\mnodla \mid \zqso) &+ \Prob(\msubdla \mid \zqso)\\
      &+ \Prob(\mdla \mid \zqso) = 1.
   \end{split}
\end{equation}

The scaling factor between the {\dla} prior and sub-{\dla} prior should depend on our prior probability density of the column density of the absorbers.
Here we assume the density of sub-{\dla} $\lognhi$ is an uniform density with a finite range of $\lognhi \in [19.5, 20]$.
We believe there are more sub-{\dla}s than {\dla}s as high column density systems are generally rarer.
We thus assume the probability of finding sub-{\dla}s at a given $\lognhi$ is the same as the probability of finding {\dla}s at the most probable $\lognhi$, which is:
\begin{equation}
   \begin{split}
      p(\lognhi &= N \mid \{\mdla, \msubdla\} ) = \\
       &\alpha q(N \mid \mdla) \indicator_{(20, 23)}(N) \\
      +& \alpha \max{( q(N \mid \mdla))} \indicator_{(19.5, 20)}(N)\\
      +& (1 - \alpha) \mathcal{U}[19.5, 23].
   \end{split}
\end{equation}
Since $q(N \mid \mdla)$ has a simple quadratic functional form, we can solve the maximum value analytically,
which is $\max{(q(N \mid \mdla))} \simeq q(N = 20.03 \mid \mdla)$.

We thus can use our prior knowledge about the logarithm of column densities for different absorbers to rescale model priors:
\begin{equation}
   \Prob(\msubdla \mid \zqso)
   = \frac{Z_\textrm{sub}}{Z_\textrm{DLA}}
   \Prob(\mdla \mid \zqso),
\end{equation}
where the scaling factor is:
\begin{equation}
   \frac{Z_\textrm{sub}}{Z_\textrm{DLA}}
   = \frac{
      \int_{19.5}^{20} p( N \mid \{\mdla, \msubdla\} ) \dd N
   }{
      \int_{20}^{23} p(N \mid \{\mdla, \msubdla\} ) \dd N
   },
\end{equation}
which is the odds of finding absorbers in the range of $\lognhi \in [19.5, 20]$ compared to finding absorbers in $\lognhi \in [20, 23]$.
Note that we will treat the model posteriors of the sub-{\dla} model as part of the non-detections of {\dla}s in the following analysis sections.


\section{Catalogue}
The original parameter prior in \cite{Garnett17} is uniformly distributed in $\zdla$ between the Lyman limit ($\lambdarest = 911.76$ \AA) and the Ly$\alpha$ emission of the quasar.
In \cite{Bird17}, we chose the minimum value of $\zdla$ to be at the Ly$\beta$ emission line of the quasar rest-frame (instead of the Lyman limit) to avoid the region containing unmodelled Ly$\beta$ forest. The primary reason for this was that the original absorption noise model did not include Ly$\beta$ absorption.
With the updated model from Eq.~\ref{eq:absorption_noise} we are able to model this absorption. Hence, for our new public catalogue, we sample $\zdla$ to be from $\Ly{\infty}$ to $\Ly{\alpha}$ in the quasar rest-frame and for the convenience of future investigators our public catalogue contains {\dla}s throughout the whole available spectrum, including $\Ly{\beta}$ to $\Ly{\infty}$. There is still some contamination in the blue end of high redshift spectra from the $\Ly\beta$ forest and occasional Lyman breaks from a misestimated quasar redshift. In practice we shall see that the contamination is not severe except for $\zdla > 3.75$. However, in the interest of obtaining as reliable {\dla} statistics as possible, when computing population statistics we consider only 3\,000~{\AA} redward of \Ly$\beta$ to 3\,000~{\AA} blueward of \Ly$\alpha$ in the quasar rest frame.

In this paper, we computed the posterior probability of $\mnodla$ to $\mkdla$ models.
For each spectrum, the catalogue includes:
\begin{itemize}
   \item The range of redshift {\dla} searched $[z_\textrm{min}, z_\textrm{max}]$,
   \item The log model priors from $\log \Prob(\mnodla \mid \zqso)$,  $\log \Prob(\msubdla \mid \zqso)$, to $\log \Prob(\{\mdla{_{(i)}}\}_{i=1}^{k} \mid \zqso)$,
   \item The log model evidence $\log p(\yvec \mid \lambdavec, \nuvec, \zqso, \mathcal{M})$, for each model we considered,
   \item The model posterior $\Prob(\mathcal{M} \mid \Data, \zqso)$, for each model we considered,
   \item The probability of having {\dla}s $\Prob(\{ \mdla \} \mid \Data, \zqso)$,
   \item The probability of having zero {\dla}s $\Prob(\mnodla \mid \Data, \zqso)$,
   \item The sample log likelihoods $\log p(\yvec \mid \lambdavec, \nuvec, \zqso, \{{\zdla}_{(i)}\}_{i=1}^{k}, \{\lognhi{_{(i)}}\}_{i=1}^{k}, {\mdla}_{(k)})$ for all {\dla} models we considered, and
   \item The maximum a posteriori ({\mapval}) values of all {\dla} models we considered.
\end{itemize}
The full catalogue will be available alongside the paper: \url{http://tiny.cc/multidla_catalog_gp_dr12q}.
The code to reproduce the entire catalogue will be posted in \url{https://github.com/rmgarnett/gp_dla_detection/tree/master/multi_dlas}.


\subsection{Running Time}
We ran our multi-{\dla} code on {\ucr}'s High-Performance Computing Center ({\hpcc}) and Amazon Elastic Compute Cloud ({\ectwo}).
The computation of model posteriors of $\mnodla$, $\msubdla$, $\{ \mdla{_{(i)}} \}_{i=1}^4$ takes 7-11 seconds per spectrum on a 32-core node in {\hpcc} and 3-5 seconds on a 48-core machine in {\ectwo}.
For each spectrum, we have to compute $10\,000 * 5 + 1$ log likelihoods in the form of Eq.~\ref{eq:model_evidence_null}.
If we scale the sample size from $N = 10\,000$ to $100\,000$, it costs 38-52 seconds on a 32-core node in {\hpcc}.

\section{Example spectra}
Here we show a few examples of the fitted {\gp} priors, both to compare our method to others and to aid the reader in understanding concretely how our method works.

\begin{figure*}
   \includegraphics[width=2\columnwidth]{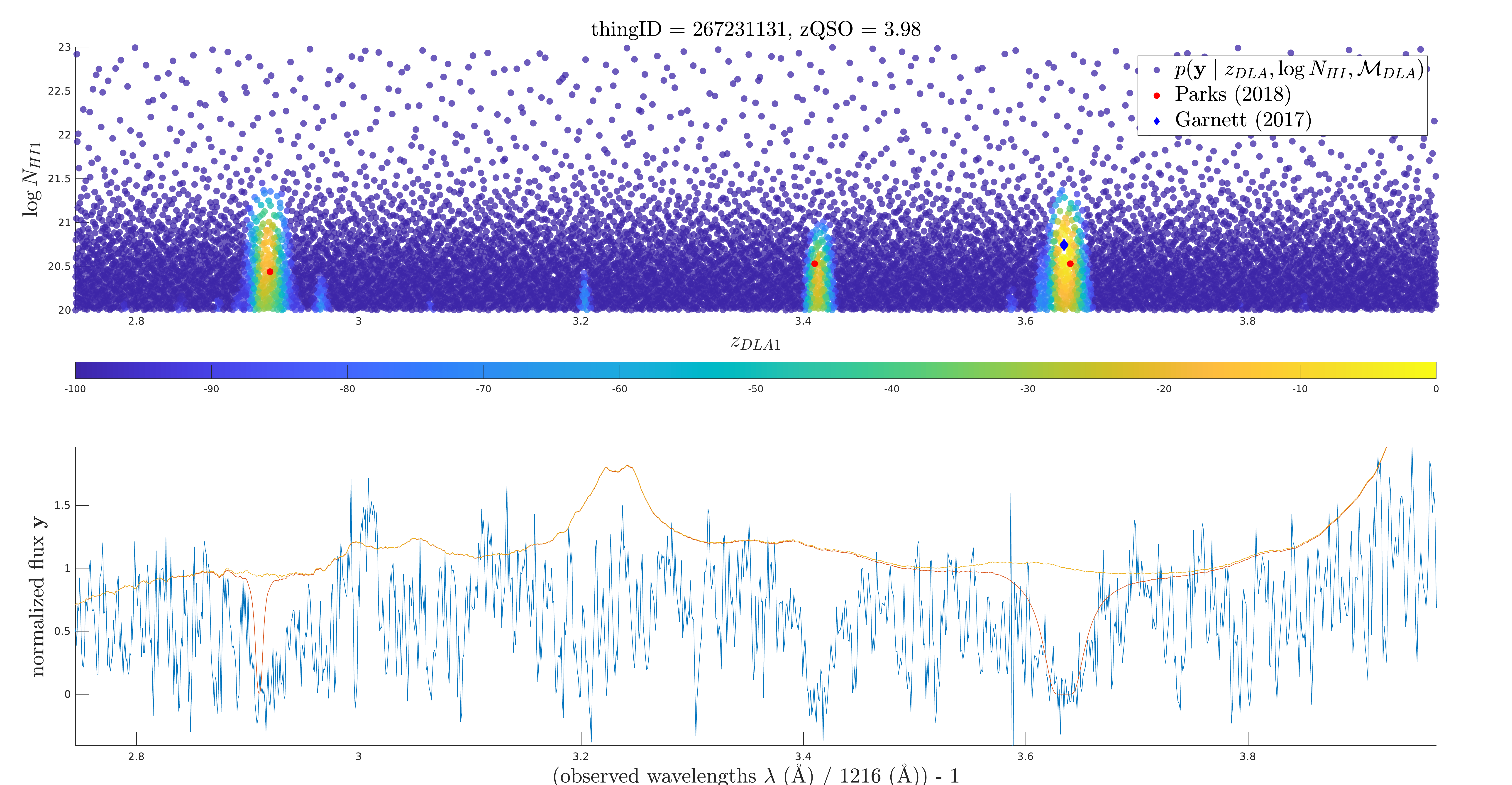}
   \caption{
      An example of finding {\dla}s using \protect\cite{Garnett17}'s model.
      Here we use the single-{\dla} per spectrum version of Garnett's model.
      \textbf{Upper:} sample likelihoods $p(\yvec \mid \theta, \mdla)$ in the parameter space $\theta = (\zdla, \lognhi)$.
      Red dots show the {\dla}s predicted by \protect\cite{Parks18}, and the blue squares show the maximum a posteriori ({\mapval}) prediction of the \protect\cite{Garnett17}.
      \textbf{Bottom:} the observed spectrum (blue), the null model {\gp} prior (orange), and the {\dla} model {\gp} prior (Red).
      So that the upper and bottom panels have the same x-axis,
      we rescale the observed wavelength to absorber redshift.
      }
   \label{fig:mu_original_lya}
\end{figure*}

\begin{figure*}
   \includegraphics[width=2\columnwidth]{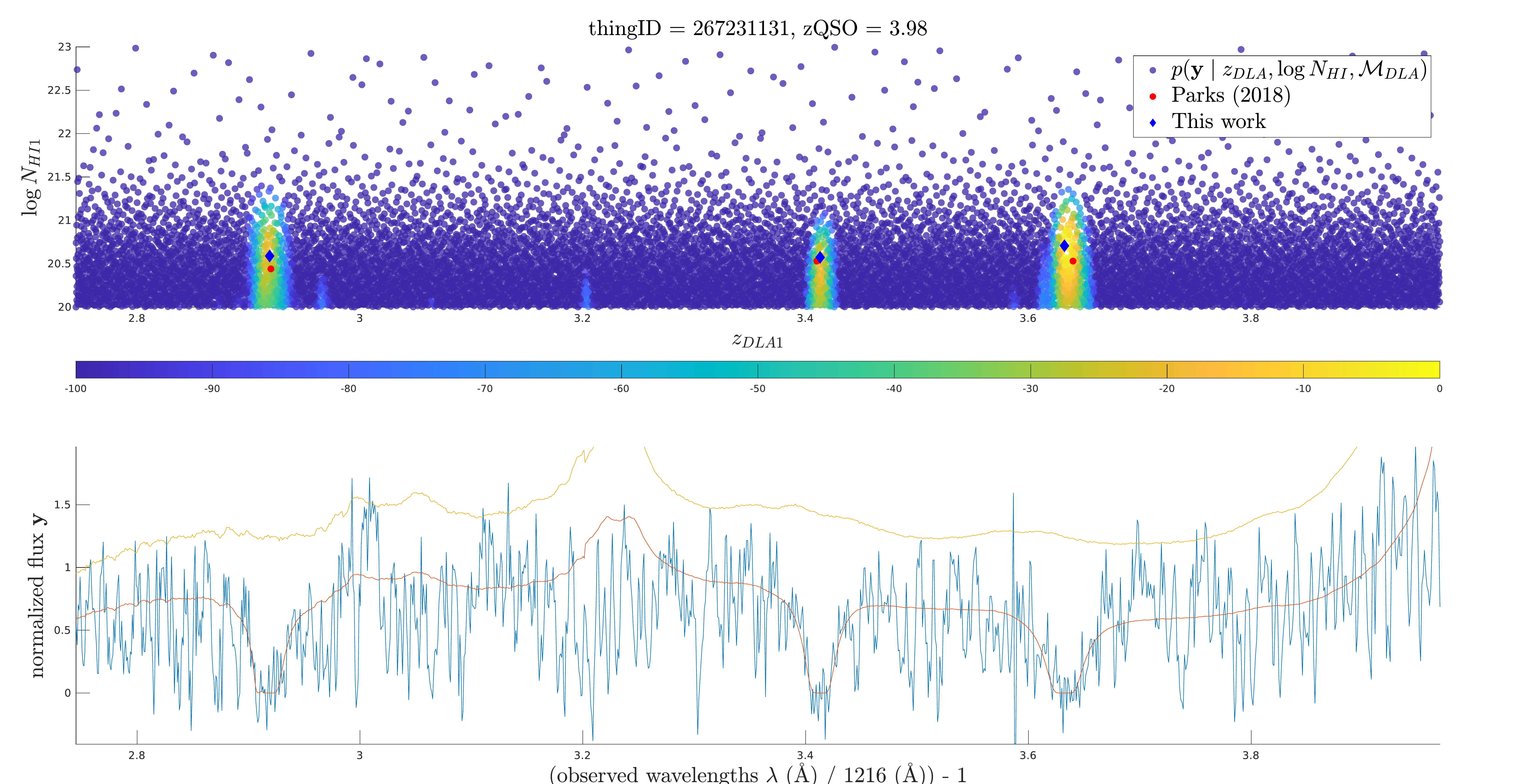}
   \caption{
      The same spectrum as Figure~\ref{fig:mu_original_lya}, but using the multi-{\dla} model reported in this paper.
      \textbf{Upper:} sample likelihoods $p(\yvec \mid \theta, \mdla)$ in the parameter space of the $\mdla{_{(1)}}$, with $\theta = (\zdla, \lognhi)$.
      \textbf{Bottom:} the observed spectrum (blue), the null model {\gp} prior before the suppression of effective optical depth (orange), and the multi-{\dla} {\gp} prior (Red).
      The orange curve is slightly higher than the one in Figure~\ref{fig:mu_original_lya} because we try to model the mean spectrum before the forest.
      However, the DLA quasar model (red curve) matches the level of the observed mean flux better than Figure~\ref{fig:mu_original_lya} due to the inclusion of a term for the effective optical depth of the \Lya~forest.
      }
   \label{fig:mu_lya1pz}
\end{figure*}

Figure~\ref{fig:mu_lya1pz} shows an example where our new code detects three {\dla}s in a single spectrum, while the older model detected only one {\dla} as shown in Figure~\ref{fig:mu_original_lya}. Because the mean quasar model includes a redshift dependent term corresponding to intervening absorbers, our new mean model can now fit the mean observed quasar spectrum better.
Although we show the sample likelihoods in the $\mdla{_{(1)}}$ parameter space, our current code finds these three {\dla}s in the six dimensional parameter space $( \zdla{_{(i)}}, \lognhi{_{(i)}} )_{i=1}^{3}$.

\begin{figure*}
   \includegraphics[width=2\columnwidth]{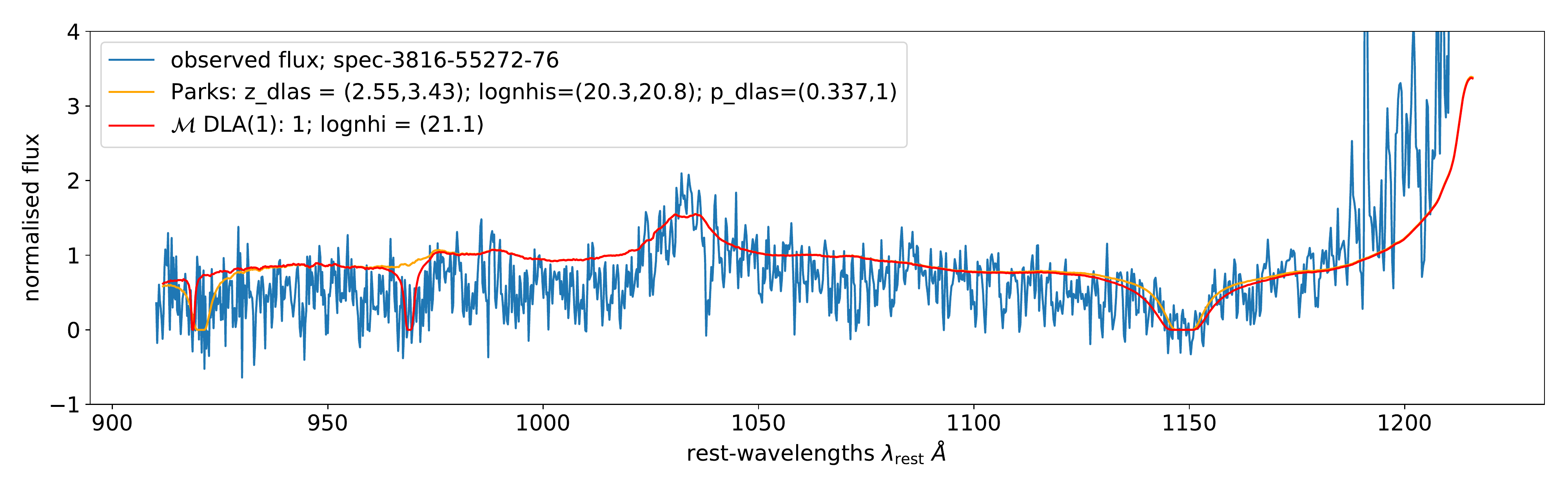}
   \caption{
      \textbf{Blue:} the normalised observed flux. The spectral ID represents \texttt{spec-plate-mjd-fiber\_id}.
      \textbf{Yellow:} Parks' predictions on top of our null model.
      Our model predicts only one {\dla} while the {\cnn} model in \protect\cite{Parks18} predicts two {\dla}s.
      One of the {\dla}s predicted by \protect\cite{Parks18} is coincident with the \Ly$\gamma$ absorption from our predicted {\dla}.
      \texttt{z\_dla} corresponds to the {\dla} redshifts reported in Parks' catalogue, and \texttt{lognhi} corresponds to the column density estimations of Parks' catalogue. \texttt{p\_dla} is the \texttt{dla\_confidence} reported in Parks.
      \textbf{Red:} Our current model with the highest model posterior and the {\mapval}s of column densities.
      In this spectrum, we show that it is crucial to include \Ly$\beta$ and \Ly$\gamma$ absorption from the {\dla} in the {\dla} profile.
      It not only helps to localize the {\dla},
      but it also predicts $\nhi$ more accurately using information from the \Ly$\beta$ region.
      The blue line shows the observed flux,
      the red curve is our multi-{\dla} {\gp} prior,
      and the orange curve shows the predicted {\dla}s from \protect\cite{Parks18} subtracted from our mean model.
   }
   \label{fig:example_two}
\end{figure*}

\begin{figure*}
   \includegraphics[width=2\columnwidth]{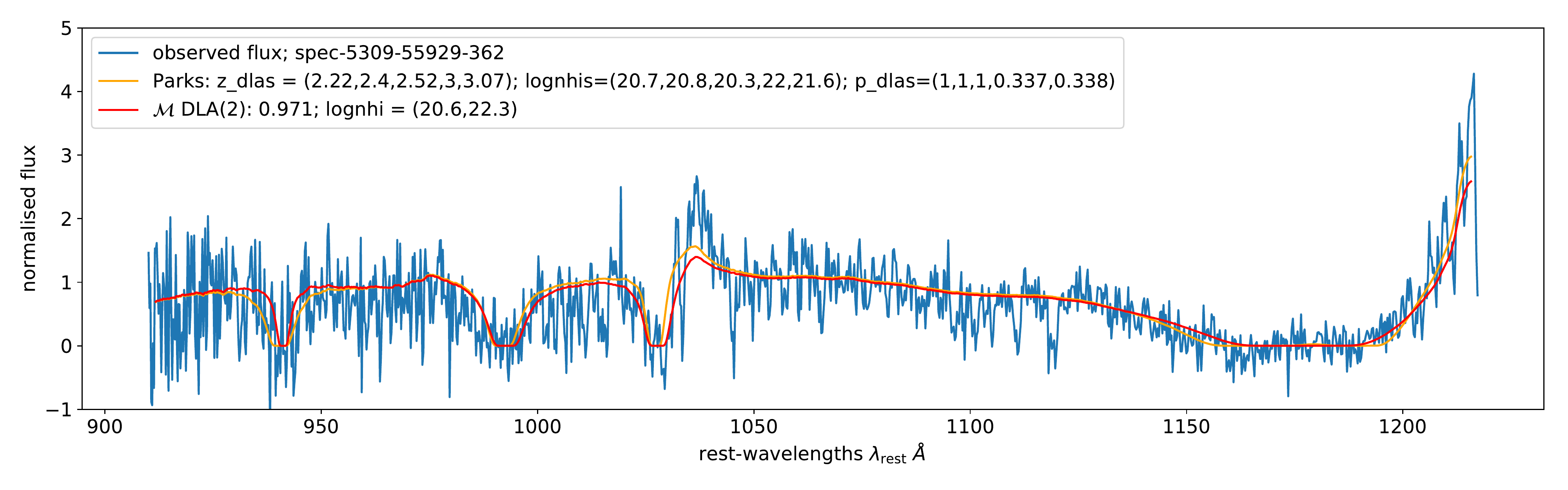}
   \caption{A spectrum in which we detect two {\dla}s.
   \textbf{Blue:} Normalised flux.
   \textbf{Red:} {\gp} mean model with two intervening {\dla}s.
   \textbf{Yellow:} The predictions from Parks' catalogue.
   \textbf{Pink:} The {\mapval} prediction of \protect\cite{Garnett17}
   on top of the {\gp} mean model without mean flux suppression.
   The model posterior from \protect\cite{Garnett17} is listed in the legend ($1$) with the {\mapval} value of $\lognhi$.
   The column density estimate for the {\dla} near $\lambdarest = 1\,025 ${\AA} has large uncertainty (see Figure~\ref{fig:example_referee_subdlas_likelihoods}). It is thus possible that this {\dla} could be a sub-{\dla}, as preferred by \protect\cite{Parks18}.
   }
   \label{fig:example_two_dlas}
\end{figure*}

In Figure~\ref{fig:example_two} we show a representative sample of a very common case in our $\mdla{_{(1)}}$ model.
The red curve represents our {\gp} prior on the given spectrum, and the orange curve is the curve with fitted {\dla}s provided by the {\cnn} model presented in \cite{Parks18}\footnote{
   We used the version of \cite{Parks18}'s catalogue listed in the published paper and found on Google Drive at \url{https://tinyurl.com/cnn-dlas}.}.
We found \cite{Parks18} underestimated the column densities of the underlying {\dla}s in the spectra due to not modelling {\Lyb} and Lyman-$\gamma$ absorption in {\dla}s, while the predictions of $\nhi$ in our model are more robust since the predicted $\nhi$ is constrained by $\alpha$, $\beta$, and $\gamma$ absorption. In the spectrum, {\Lyb} absorption is clearly visible (although noisy). In Figure~\ref{fig:example_two}, \cite{Parks18} has actually mistaken the \Ly$\gamma$ absorption line of the {\dla} for another, weaker, {\dla}.
This demonstrates again the necessity of including other Lyman-series members in the modelling steps.
Since \cite{Parks18} broke down each spectrum into pieces during the training and testing phases, it is impossible for the {\cnn} to use knowledge about other Lyman series lines associated with the {\dla}s.
Another example, from a spectrum where we detect $2$ {\dla}s and the {\cnn} detects $4$ (although at low significance) is shown in Figure~\ref{fig:example_two_dlas}. Here the {\cnn} has mistaken both the \Ly$\beta$ and \Ly$\gamma$ absorption associated with the large {\dla} at $z\sim 3$ (near the quasar rest frame) for separate {\dla}s at $z=2.4$ and $z = 2.22$ respectively. The large {\dla} at $z \sim 3$ has been split into two of reduced column density and reduced confidence. The {\cnn} has also missed the second genuine $\dla$ at a rest-frame wavelength of $1025$\AA, presumably due to the proximity of an emission line. Our code, able to model the higher order Lyman lines, has used the information contained within them to correctly classify this spectrum as containing two {\dla}s.

\begin{figure*}
   \includegraphics[width=2\columnwidth]{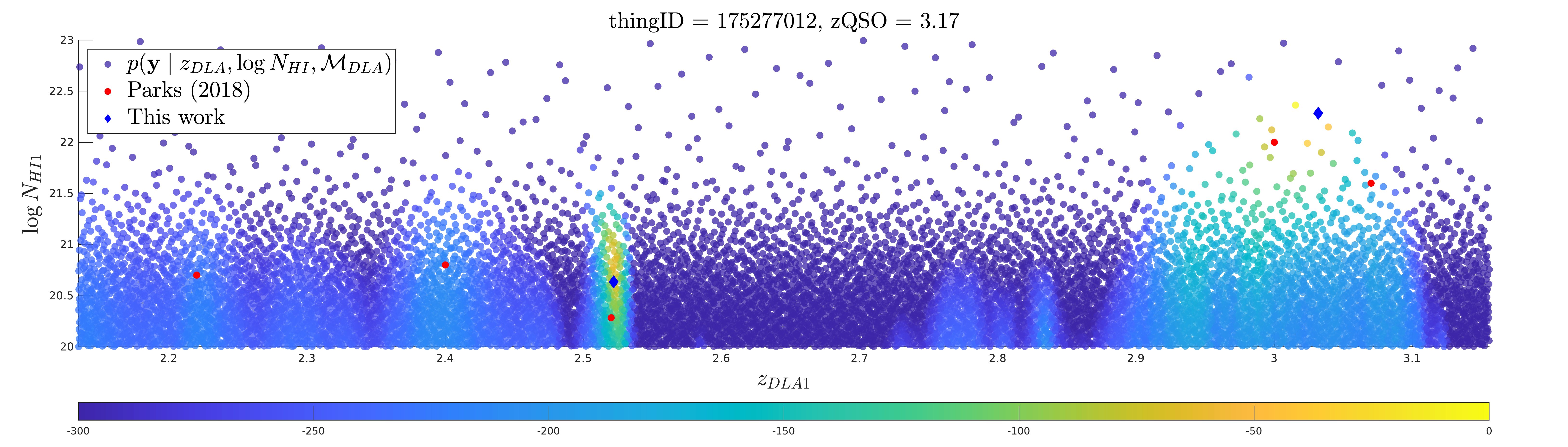}
   \caption{
   The log sample likelihoods for the {\dla} model of the spectrum shown in Figure~\ref{fig:example_two_dlas}, normalised to range from $-\infty$ to $0$.
   The {\dla} at $\zdla \sim 2.52$ could be a sub-{\dla} (as preferred by \protect\cite{Parks18}), as the {$\lognhi$} estimate is uncertain.
   However, we found that the 2-{\dla} model posterior $\log{p(\mdla{_{(2)}} \mid \yvec, \lambdavec, \nuvec, \zqso )} = -638$ is still higher than the model posterior from combining 1-{\dla} and 1-sub-{\dla}, which is $\log{p(\mdla{_{(1)}} + \msubdla \mid \yvec,\lambdavec, \nuvec, \zqso )} = -691.47$.
   }
   \label{fig:example_referee_subdlas_likelihoods}
\end{figure*}

\begin{figure*}
   \includegraphics[width=2\columnwidth]{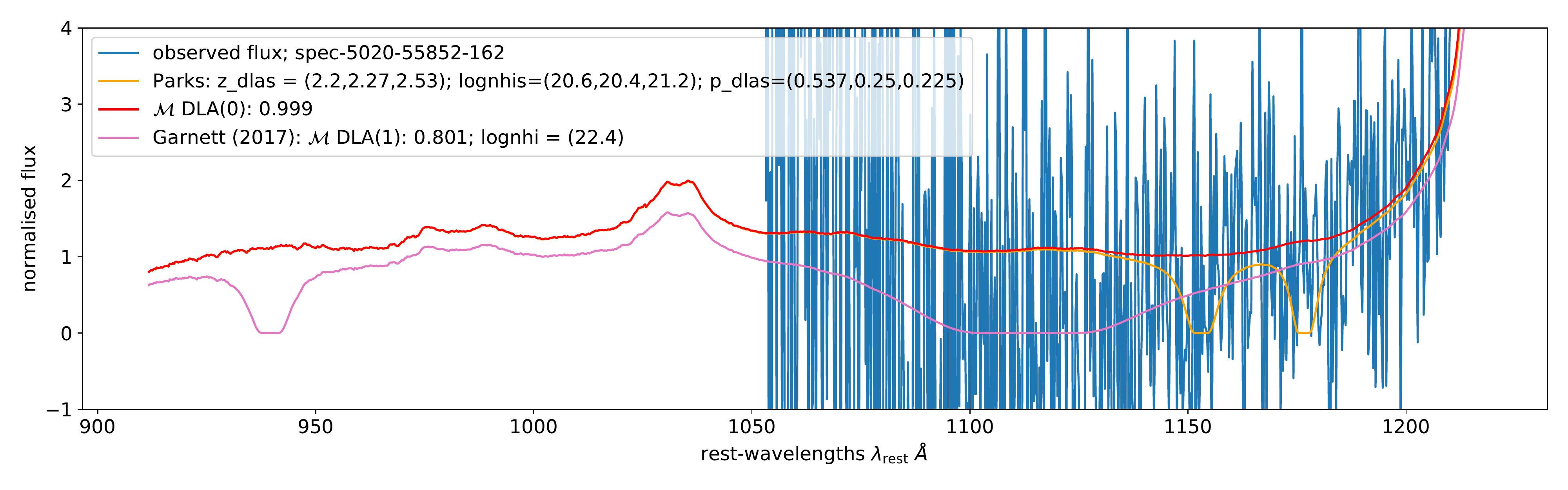}
   \caption{A noisy spectrum at $\zqso = 2.378$ fitted with a large {\dla} by \protect\cite{Garnett17}.
   \textbf{Red:} The model presented in this paper predicts no {\dla} detection in thie spectrum.
   \textbf{Pink:} The {\mapval} prediction of \protect\cite{Garnett17} on top the {\gp} mean model without the mean-flux suppression.
   \textbf{Gold:} The prediction of \protect\cite{Parks18} subtracted from our mean model. Note that \protect\cite{Parks18} also indicates a detection of a {\dla} at $\zdla = 2.53$, but outside the range of this spectrum.}
   \label{fig:example_additional_occams}
\end{figure*}

\begin{figure*}
   \includegraphics[width=2\columnwidth]{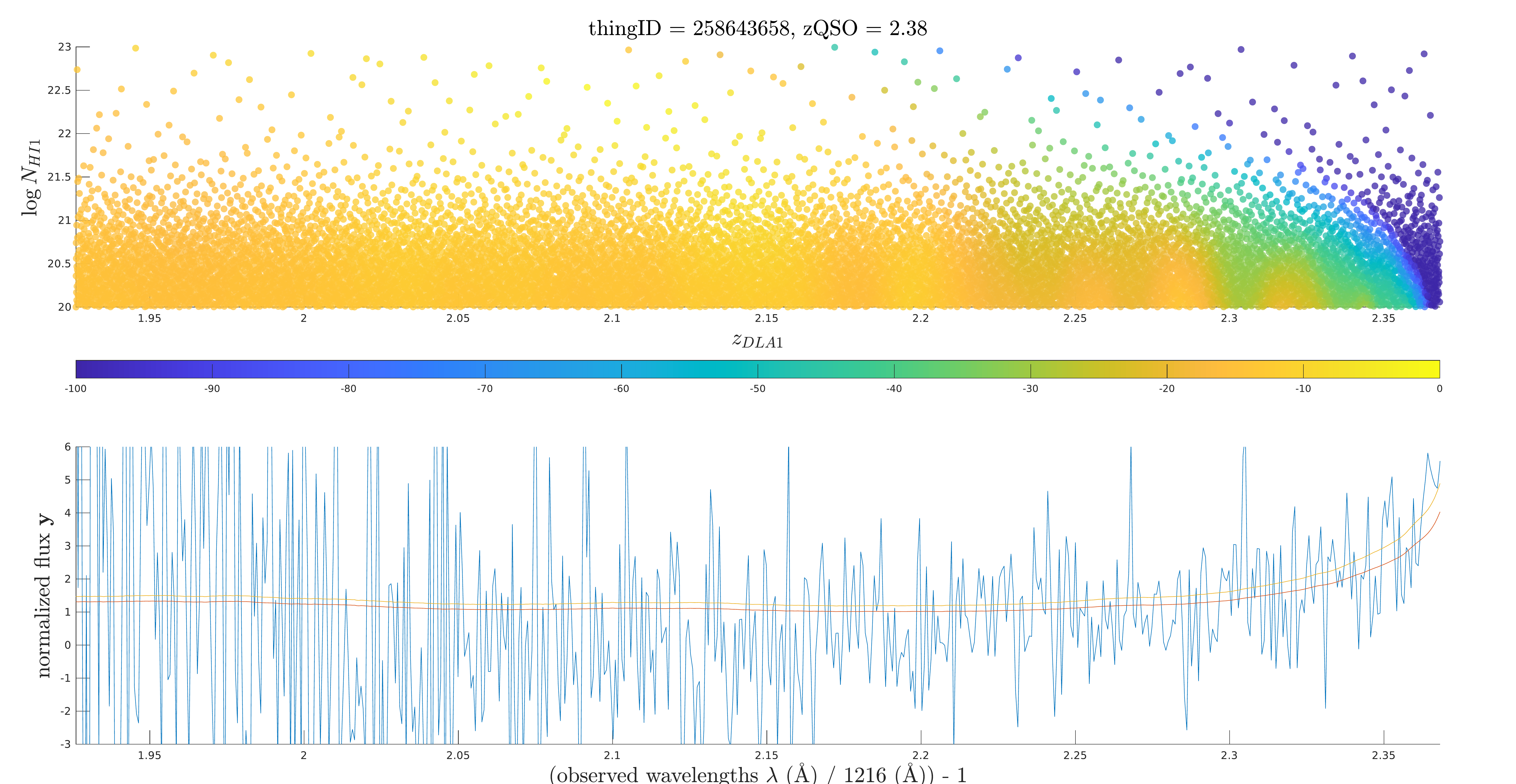}
   \caption{
   \textbf{Top:} The sample likelihoods of the spectrum shown in Figure~\ref{fig:example_additional_occams}.
   The colour bar indicates the normalised log likelihoods ranging from $-\infty$ to $0$.
   \textbf{Bottom:} The orange curve indicates the {\gp} mean model before mean-flux suppression, the red curve represents the mean model after suppression,
   and the blue line is the normalised flux of this spectrum.
   The x-axis of this spectrum is rescaled to be the same as the $\zdla$ presented in the upper panel.
   }
   \label{fig:example_additional_occams_likelihoods}
\end{figure*}

Figure~\ref{fig:example_additional_occams} shows an example which was problematic in both the models of \cite{Garnett17} and \cite{Parks18}.
This is an extremely noisy spectrum, where the length of the spectrum is not long enough for us to contain higher order \Ly-series absorption or even to see the full length of the putative {\Lya} absorption. By eye, distinguishing a {\dla} from the noise is challenging. If we examine the sample likelihoods from our model (shown in Figure~\ref{fig:example_additional_occams_likelihoods}), we see that the {\dla} posterior probability is spread over the whole of parameter space; in other words, all models are a poor fit for this noise-dominated spectrum. The model selection is thus really comparing the likelihood function on the basis of how much parametric freedom it has. After implementing the additional Occam's razor factor between the null model and parameterised models ({\dla}s and sub-{\dla}s) described in Section~\ref{sec:additional_occams_razor},
we found that the large {\dla} fitted to the noisy short spectrum by \cite{Garnett17} was no longer preferred. This indicates that our Occam's razor penalty is effective. As shown in Figure~\ref{fig:omega_dla_main}, $\omegadla$ at low redshifts is lower than the measurements in \cite{Bird17}, indicating that this class of error is common enough to have a measurable effect on the column density function. We checked that the addition of the Occam's razor penalty, $\omegadla$ is insensitive to the noise threshold used when selecting the spectra for our sample.

There are still some very high redshift quasars ($\zqso \gtrsim 5$) where our code clearly detects too many {\dla}s in a single spectrum, even at low redshift. We exclude these spectra from our population statistics. At high redshift the \Lya~forest absorption is so strong as to render the observed flux close to zero.
We thus cannot easily distinguish between the null model and the {\dla} models. It is also possible that at high redshifts the mean flux of the forest is substantially different from the \cite{Kim07} model we assume, and that this biases the fit. Finally, there are few such spectra, and so we cannot rule out the possibility that covariance of their emission spectra differs quantitatively from lower redshift quasars.

\section{Analysis of the results}
In this section, we present results from our classification pipeline, and we also present the statistical properties ({\cddf}, line densities $\dd N/\dd X$, and total column densities $\omegadla$) of the {\dla}s detected in our catalogue.


\subsection{ROC analysis}
\label{subsec:roc_analysis}
\begin{figure}
   \centering
   \includegraphics[width=1\columnwidth]{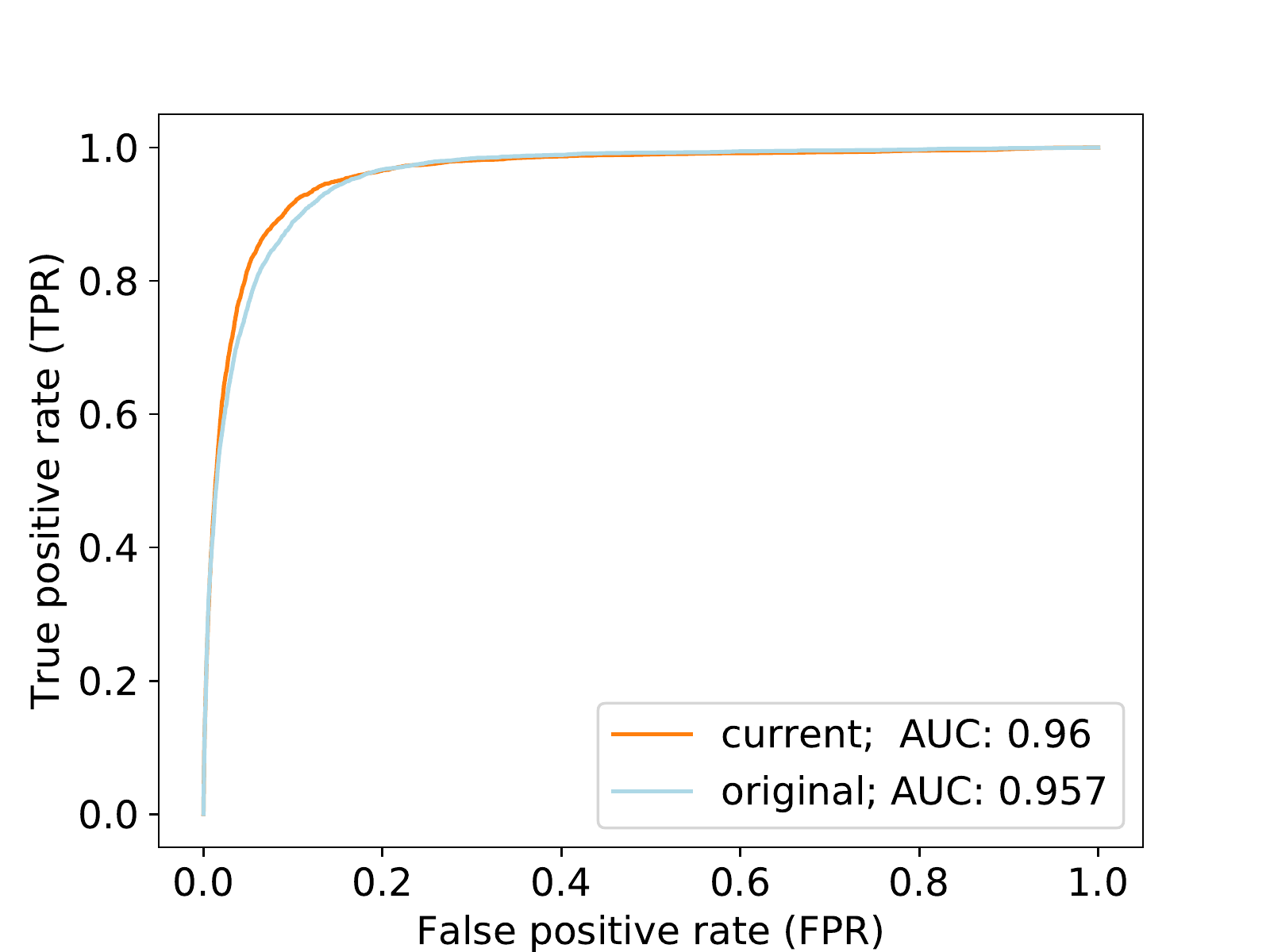}
   \caption{ The {\roc} plot made by ranking the sightlines in {\boss} {\dr}9 samples using the log posterior odds of containing at least one {\dla}.
   Ground truths are from the {\dr}9 concordance catalogue.
   The orange curve shows the {\roc} plot of our current multi-{\dla} model,
   and the blue curve is derived from \protect\cite{Garnett17}.
   In this plot we consider only the model containing at least one {\dla} $p(\{\mdla\} \mid \Data)$, rather than the multiple {\dla}s models, as the concordance catalogue contains only one {\dla} per spectrum.}
   \label{fig:ROC_DR9}
\end{figure}

To evaluate how well our multi-{\dla} classification reproduces earlier results,
we rank our {\dla} detections using the log posterior odds between the {\dla} model (summing up all possible {\dla} models $\midlak$) and the null model:
\begin{equation}
   \begin{split}
      \log &(\textrm{odds}) =\\
      &\log{\Prob(\{\mdla\} \mid \Data, \zqso )} - \log{ \Prob( \mnodla \mid \Data, \zqso ) },
   \end{split}
\end{equation}
where the ranking is over all sightlines.
From the top of the ranked list based on the log posterior odds,
we calculate the true positive rate and false positive rate for each rank:
\begin{equation}
   \begin{split}
      \textrm{TPR} &= \frac{\textrm{TP}}{ \textrm{TP} + \textrm{FN}};\\
      \textrm{FPR} &= \frac{\textrm{FP}}{ \textrm{FP} + \textrm{TN} }.
   \end{split}
\end{equation}
The true positive rate is the fraction of sightlines where we detect {\dla}s (ordered by their rank) divided by the number of sightlines with {\dla}s detected by earlier catalogues. The false positive rate is the number of detections of {\dla}s divided by the number of sightlines where earlier catalogues did not detect {\dla}s. In Figure~\ref{fig:ROC_DR9} we show the {\tpr} and {\fpr} in a receiver-operating characteristics ({\roc}) plot to show how well our classification performs. We have compared to the concordance {\dla} catalogue \citep{Lee2013} in the hope that it approximates ground truth, there being no completely reliable {\dla} catalogue.


\begin{figure}
   \centering
   \includegraphics[width=1\columnwidth]{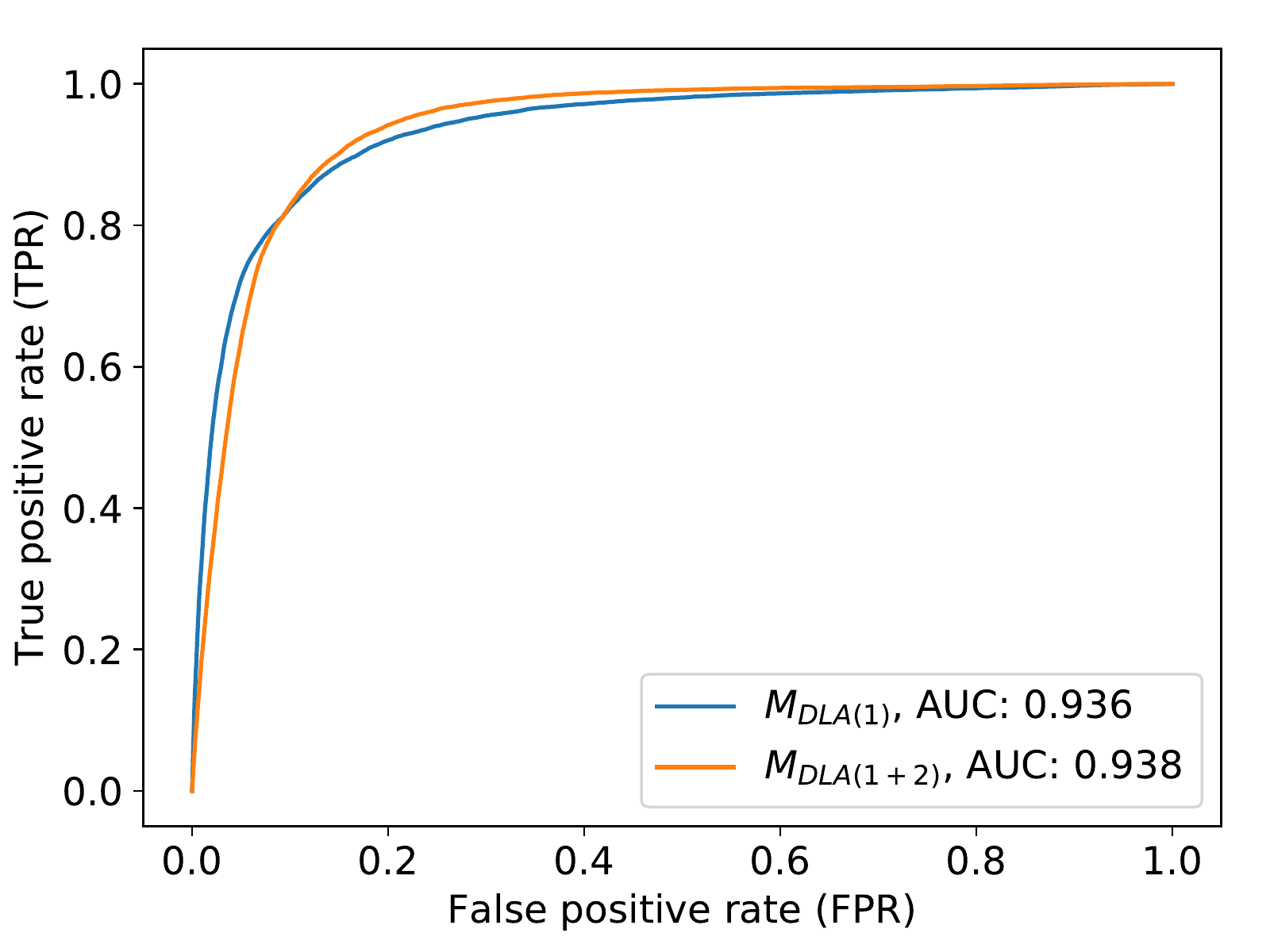}
   \caption{
      The {\roc} plot for sightlines with one and two {\dla} detections,
      by using the catalogue of \protect\cite{Parks18} (with \texttt{dla\_confidence} $> 0.98$) as ground truth.
   }
   \label{fig:ROC_Parks}
\end{figure}

We also want to know how well our pipeline can identify the number of {\dla}s in each spectrum.
The {\dr}9 concordance catalogue does not count multiple {\dla} spectra, and so we compare our multi-{\dla} detections
to the catalogue published by \cite{Parks18}.
Each {\dla} detected in \cite{Parks18} comes with a measurement of their confidence of detection (\texttt{dla\_confidence} or $p_\textrm{DLA}^\textrm{Parks}$) and a MAP redshift and column density estimate. We compare our multiple \dla catalogue to those spectra with $p_\textrm{DLA}^\textrm{Parks} > 0.98$.
The resulting {\roc} plot is shown in Figure~\ref{fig:ROC_Parks}.
We count a maximum of 2 {\dla}s in each spectrum: 3 or more {\dla}s in a single sightline are extremely rare and do not provide a large enough sample for an {\roc} plot. Parks' catalogue is not a priori more reliable than ours, especially in spectra with multiple {\dla}s,
but comparing the first two {\dla}s is a reasonable way to validate our method's ability to detect multiple {\dla}s.

These spectra are counted by breaking down each two-{\dla} sightline (either in Parks or our catalogue) into two single observations.
For example, if there are two {\dla}s detected in Parks and one {\dla} detected in our pipeline for an observation $\Data$,
we will assign one ground-truth detection to $p(\mdla{_{(1)}} \mid \Data)$ and assign one ground-truth detection to $p(\mnodla \mid \Data)$.
On the other hand, if there is only one {\dla} detected in Parks and two {\dla}s detected in our pipeline,
we will assign one ground-truth detection to $p(\mdla{_{(2)}} \mid \Data)$ and one ground-truth non-detection to $p(\mdla{_{(2)}} \mid \Data)$.

\begin{figure*}
   \centering
   \includegraphics[width=2\columnwidth]{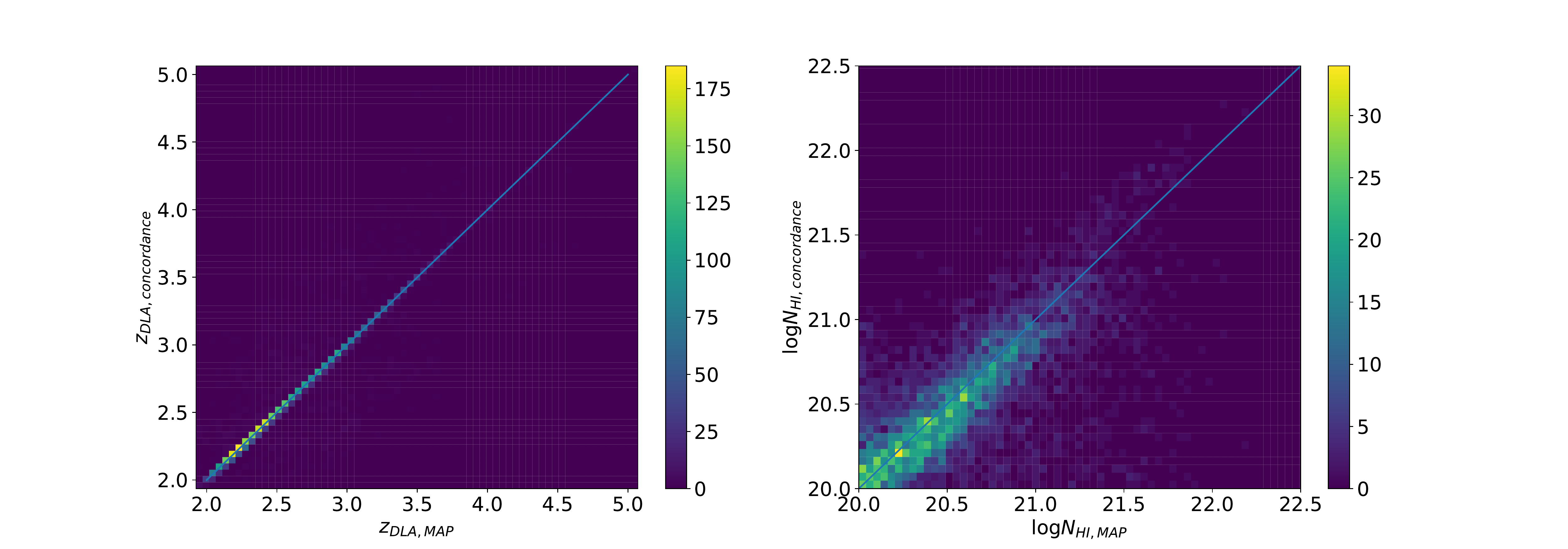}
   \caption{
      The {\mapval} estimates of the {\dla} parameters $\theta = (\zdla, \lognhi)$ for {\dla}s detected by our model in spectra observed by {\sdss} {\dr}9, compared to the values reported in the concordance catalogue.
      The straight line indicates a perfect fit.
      Note that the concordance $\lognhi$ values are not ground truth, so the scatter in column density predictions was expected.  
      }
   \label{fig:map_concordance}
\end{figure*}

In Figure~\ref{fig:map_concordance}, we also analyse the \textit{maximum a posteriori} ({\mapval}) estimate of the parameters $(\zdla, \lognhi)$ by comparing with the reported values in {\dr}9 concordance {\dla} catalogue.
The median difference between these two is $-2.2 \times 10^{-4} \, (-66.6 \, \kms)$ and the interquartile range is $2.2\times 10^{-3} \, (662 \, \kms)$.
For the log column density estimate, 
the median difference is $0.040$, and the interquartile range is $0.26$.
The medians and interquartile ranges of the {\mapval} estimate are very similar to the values reported in \cite{Garnett17} with the median of {$\zdla$} slightly smaller and the median of {$\lognhi$} slightly larger.
Note that the {\dr}9 concordance catalogue is not the ground truth, so small variations in comparison to \cite{Garnett17} can be considered to be negligible.
As shown in Figure~\ref{fig:map_concordance}, both histograms are roughly diagonal, although the scatter in column density {\mapval} is large.
Note that our {\dla}-detection procedure is designed to evaluate the model evidence across all of parameter space: a single sample {\mapval} cannot convey the full posterior probability distribution. In Section~\ref{subsec:cddf_analysis}, we thus describe a procedure to propagate the posterior density in the parameter space directly to column density statistics.

\subsection{CDDF analysis}
\label{subsec:cddf_analysis}
We follow \cite{Bird17} in calculating the statistical properties of the modified {\dla} catalogue presented in this paper.
We summarise the properties of {\dla}s using the averaged binned column density distribution function ({\cddf}), the incident probability of {\dla}s ($\dd N/\dd X$), and the averaged matter density as a function of redshift ($\Omega_\textrm{DLA}(z)$).

To plot these summary statistics, we need to convert the probabilistic detections in the catalogue to the expected average number of {\dla}s and their corresponding variances.
We first describe how we compute the expected number of {\dla}s in a given column density and redshift bin.
Next, we show how we derive the {\cddf}, $\dd N/\dd X$, and $\Omega_\textrm{DLA}(z)$ from the expected number of {\dla}s.
A sample of $n$ observed spectra contains a sequence of $n$ model posteriors $p^{1}_\textrm{DLA}, p^{2}_\textrm{DLA}, ..., p^{n}_\textrm{DLA}$ defined by:
\begin{equation}
   \pidla
   = p(\{\mdla\} \mid \yvec_i, \lambdavec_i, \nuvec_i, \zqso{_i}),
\end{equation}
where $i = 1, 2, ..., n$ is the index of the spectrum, and the {\dla} model here includes all computed {\dla} models $\{ \mdla \} = \midlak $, so that $k=4$ is the maximum possible number of {\dla}s in each spectrum in our model.

Suppose the region of interest is in a specific bin $\boldTheta$, an interval in the parameter space of column density or {\dla} redshift $\boldTheta \in \{ \nhi, \zdla \}$.
To compute the posterior of having {\dla}s in each spectrum in a given bin $\boldTheta$, $\pidla( \{\mdla\} \mid \boldTheta)$, we integrate over the sample likelihoods in the bin and multiply the model posterior by the total $\pidla$ for spectrum $i$:
\begin{equation}
   \begin{split}
      \pidlabin &\propto\\
      \pidla \times &\int_{\underline{\boldTheta}}^{\overline{\boldTheta}}
      p(\yvec_i \mid \{\mdla\}, \lambdavec_i, \nuvec_i, \zqso{_i}, \theta) d\theta\,.
   \end{split}
\end{equation}
$\theta$ is either $\zdla$ or $\lognhi$ and $\theta \in \boldTheta = (\underline{\boldTheta}, \overline{\boldTheta})$.

We calculate the posterior probability of having $N$ {\dla}s by noting that the full likelihood follows the Poisson-Binomial distribution.
Consider a sequence of trials with a probability of success equal to $\pidlabin \in [0, 1]$. The probability of having $N$ {\dla}s out of a total of $n$ trials is the sum of all possible $N$ {\dla}s subsets in the whole sample:
\begin{equation}
   \begin{split}
      \textrm{Pr}(N) =\\
      \sum_{\DLA \in F_{N}} &\prod_{i \in \DLA} \pidlabin \\
      &\prod_{j \in \DLA^{c}} (1 - \pjdlabin)
      \label{eq:poission_binomial}
   \end{split}
\end{equation}
where $F_{N}$ corresponds to all subsets of $N$ integers that can be selected from the sequence $\{ 1, 2, ..., n\}$.
The above expression means we select all possible $N$ choices from the entire sample, calculate the probability of those $N$ choices having {\dla}s and multiply that by the probability of the other $n - N$ choices having no {\dla}s.
If all $\pidlabin$ are equal, the Poisson-Binomial distribution reduces to a Binomial distribution.

The above Poisson-Binomial distribution is not trivial to compute given our large sample size.
The technical details of how to evaluate Eq.~\ref{eq:poission_binomial} efficiently are described in \cite{Bird17}.
In short, we use \cite{Lecam1960}'s theorem to approximate those spectra with $\pidlabin < p_\textrm{switch} = 0.25$ by an ordinary Poisson distribution, and evaluate the remaining samples with the discrete Fourier transform \citep{Fernandez2010}.
Our catalogue contains the posteriors of samples in a given spectrum.
Combined with the above probabilistic description of the total number of {\dla}s in the entire sample, we are able to obtain not only the point estimation of $\textrm{Pr}(N)$ but also its probabilistic density interval.

We thus compute the column density distribution function in a given bin $\boldTheta = \nhi \in [\nhi, \nhi + \Delta\nhi]$ with:
\begin{equation}
   \begin{split}
      f(N) = \frac{F(N)}{\Delta N \Delta X(z)}
   \end{split}
\end{equation}
where $F(N) = \mathbb{E}(N \mid \nhi \in [\nhi, \nhi + \Delta\nhi])$ is the expected number of absorbers at a given sightline within a column density interval.
Thus, the column density distribution function ({\cddf}) $f(N)$ is the expected number of absorbers per unit column density per unit absorption distance, within a given column density bin.

The definition of absorption distance $\Delta X(z)$ is:
\begin{equation}
   X(z) = \int_0^{z} (1 + z')^2 \frac{H_0}{H(z')} \dd z',
\end{equation}
which includes the contributions of the Hubble function $H^2(z) / H_0^2 = \Omega_{\textrm{M}} (1 + z)^3 + \Omega_{\Lambda}$, with $\Omega_{\textrm{M}}$ is the matter density and $\Omega_{\Lambda}$ is the dark energy density.

The incident rate of {\dla}s $\dd N/\dd X$ is defined as:
\begin{equation}
   \frac{\dd N}{\dd X}
   = \int_{10^{20.3}}^{\infty} f(N \mid \nhi, X \in [X, X + \dd X]) \dd \nhi,
\end{equation}
which is the expected number of {\dla}s per unit absorption distance.

The total column density $\Omega_\textrm{DLA}$ is defined as:
\begin{equation}
   \begin{split}
      &\Omega_\DLA = \\
      &\frac{m_\textrm{P} H_0}{c \rho_c}
      \int_{10^{20.3}}^{\infty} \nhi f(N \mid \nhi, X \in [X, X + \dd X]) \dd \nhi,
   \end{split}
\end{equation}
where $\rho_c$ is the critical density at $z = 0$ and $m_{\textrm{P}}$ is the proton mass.

\subsection{Statistical properties of DLAs}
\label{subsec:statistical_dlas}
\begin{figure}
   \centering
   \includegraphics[width=1\columnwidth]{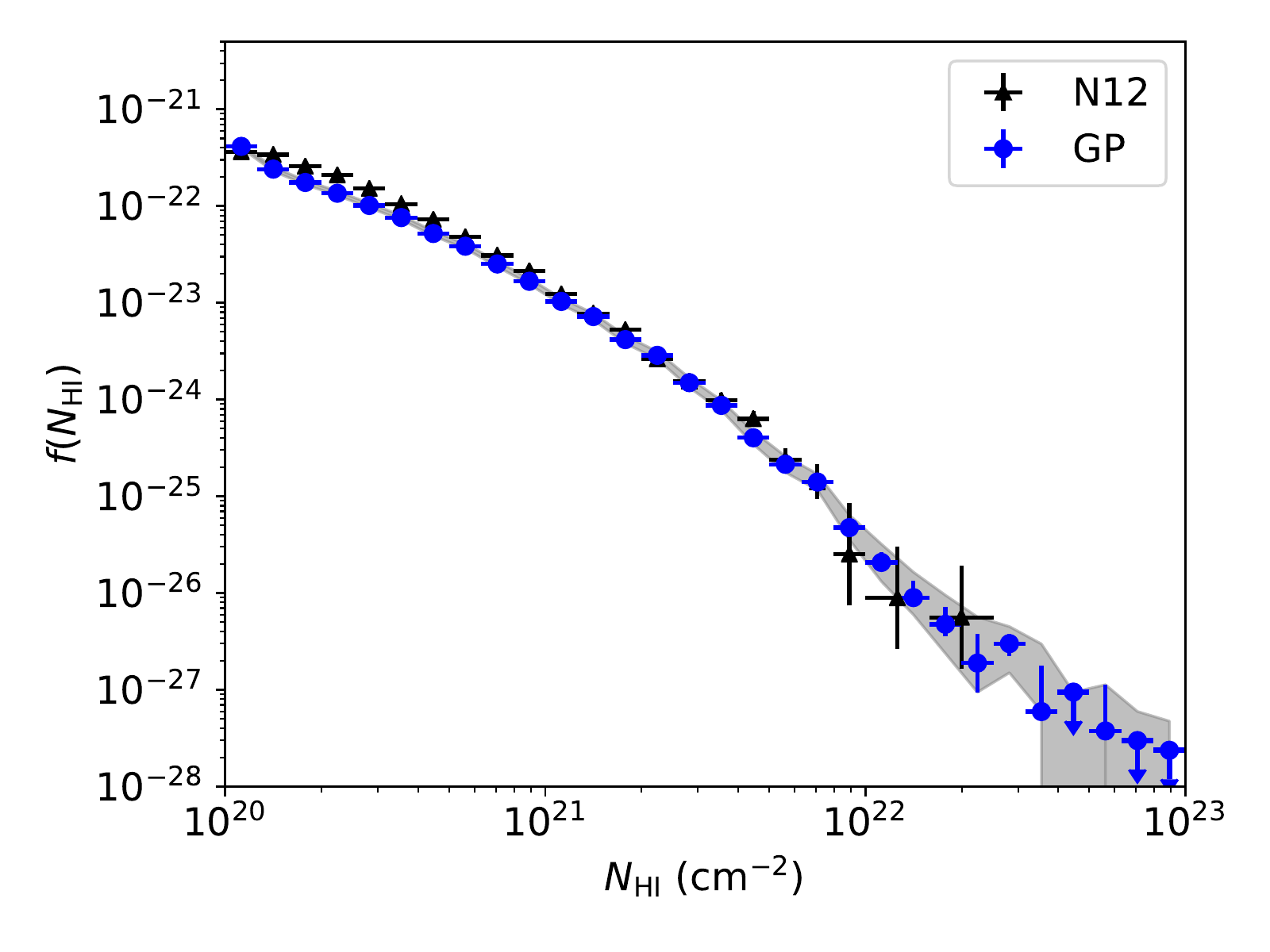}
   \caption{
      The {\cddf} based on the posterior densities for at least one {\dla} (blue, `{\gp}').
      The {\dla}s are derived from {\sdss} {\dr}12 spectra using the method presented in this paper.
      We integrate all spectral lengths with $z < 5$.
      We also plot the {\cddf} of \protect\cite{Noterdaeme12} ({\nonetwo}; black) as a comparison.
      The error bars represent the 68\% confidence limits, while the grey filled band represents the 95\% confidence limits.
      Note that our {\cddf} completely overlaps with those of {\nonetwo} for column densities in the range $10^{21}\cm^{-2}<\nhi < 10^{22}\cm^{-2}$.
   }
   \label{fig:cddf_main}
\end{figure}

\begin{figure}
   \centering
   \includegraphics[width=1\columnwidth]{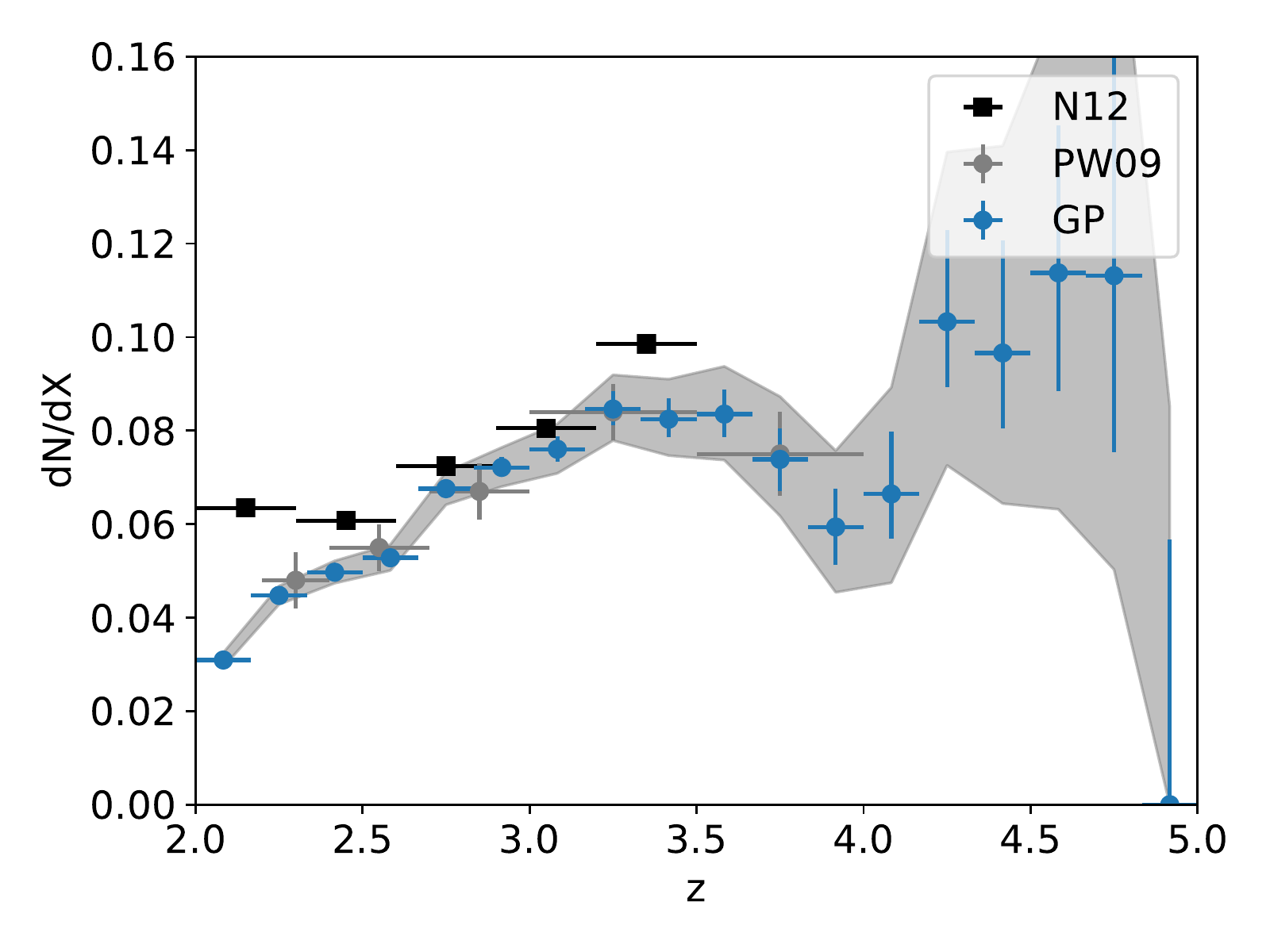}
   \caption{
      The line density of {\dla}s as a function of redshift from our {\dr}12 multi-{\dla} catalogue (blue, `{\gp}').
      We also plot the results of \protect\cite{Noterdaeme12} ({\nonetwo}; black) and \protect\cite{Prochaska2009} ({\pwzeronine}; grey).
      Note that statistical error was not computed in \protect\cite{Noterdaeme12}.
   }
   \label{fig:dndx_main}
\end{figure}

\begin{figure}
   \centering
   \includegraphics[width=1\columnwidth]{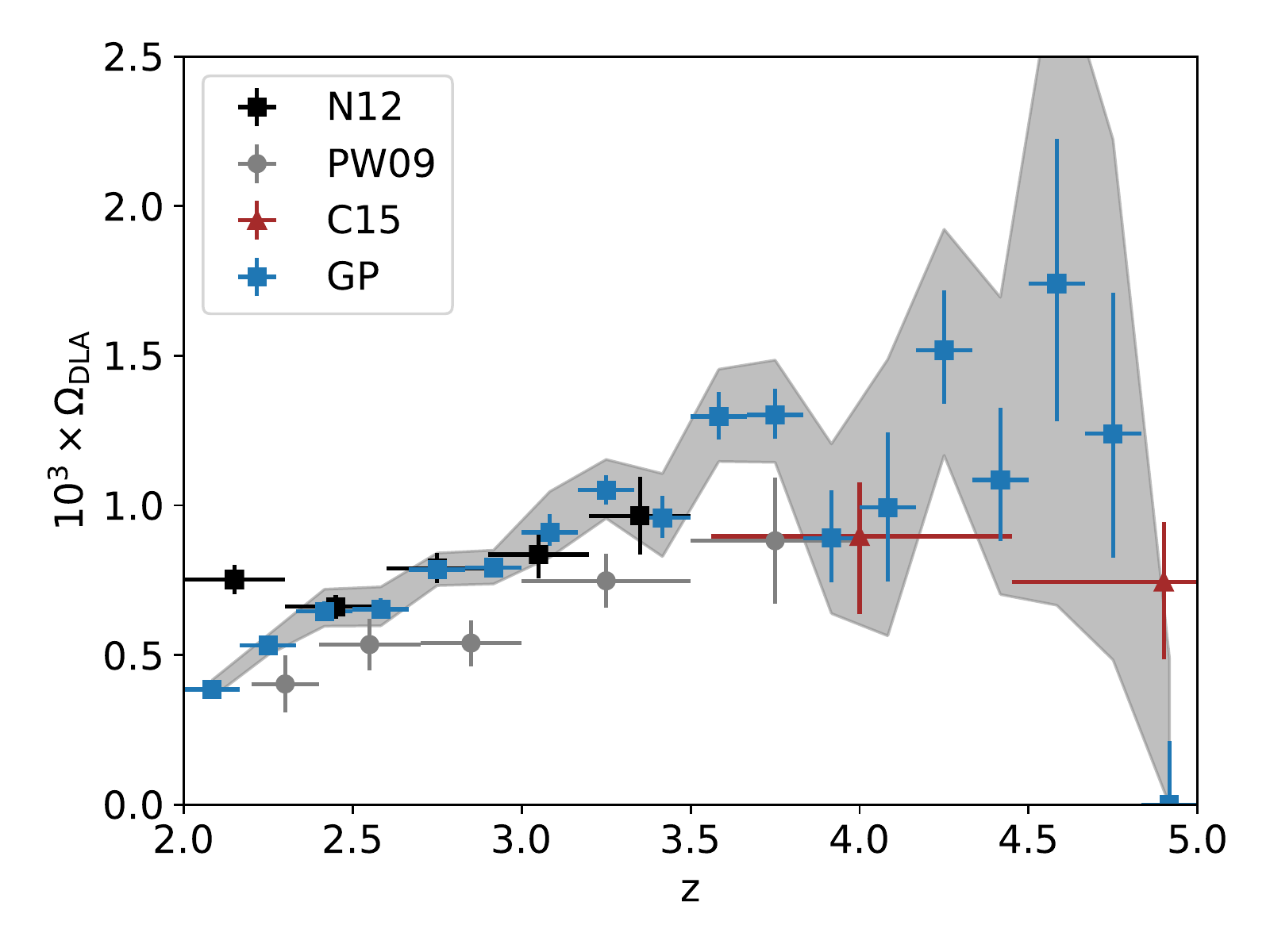}
   \caption{
      The total {\hi} density in {\dla}s, $\omegadla$, from our {\dr}12 multi-{\dla} catalogue as a function of redshift (blue, `{\gp}'),
      compared to the results of \protect\cite{Noterdaeme12} ({\nonetwo}; black), \protect\cite{Prochaska2009} ({\pwzeronine}; grey) and \protect\cite{Crighton2015} ({\conefive}; red).
   }
   \label{fig:omega_dla_main}
\end{figure}

Based on the above calculations,
we show our {\cddf} in Figure~\ref{fig:cddf_main}, $\frac{\dd N}{\dd X}$ in Figure~\ref{fig:dndx_main}, and $\omegadla$ in Figure~\ref{fig:omega_dla_main}.\footnote{The table files to reproduce Figure~\ref{fig:cddf_main} to Figure~\ref{fig:omega_dla_main} will be posted in \url{http://tiny.cc/multidla_catalog_gp_dr12q}}
Note that for determining the statistical properties of {\dla}s, we limit the samples of $\zdla$ to the range redward of the {\Lyb} in the {\qso} rest-frame, as in \cite{Bird17}.

\begin{figure}
   \centering
   \includegraphics[width=1\columnwidth]{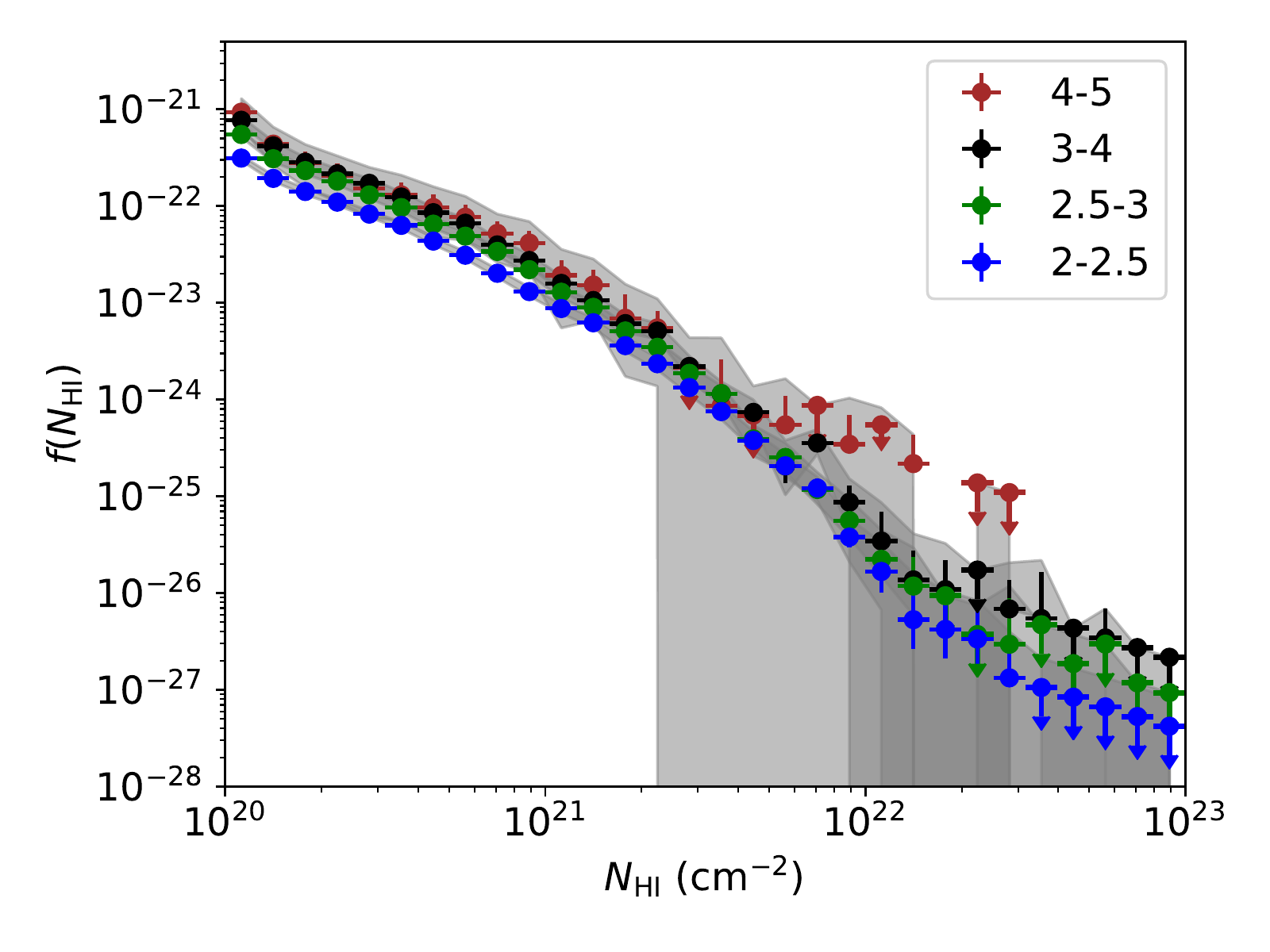}
   \caption{
      The redshift evolution (or non-evolution) of the {\cddf}.
      Labels show the absorber redshift ranges used to plot the {\cddf}s.
      In column density and redshift ranges with no detection at 68\% confidence,
      a down-pointing arrow is shown indicating the 68\% upper limit.
   }
   \label{fig:cddf_zz_main}
\end{figure}

Figure~\ref{fig:cddf_main} shows the {\cddf} from our {\dr}12 catalogue in comparison to the {\dr}9 catalogue of \cite{Noterdaeme12}.
Our {\cddf} analysis combines all spectral paths with {\qso} redshift smaller than 5, $\zdla < 5$.
The {\cddf} statistics are dominated by the low-redshift absorbers, as demonstrated in Figure~\ref{fig:cddf_zz_main}.
The error bars represent the 68\% confidence interval,
while the grey shaded area encloses the 95\% highest density region.
The {\cddf} values in Figure~\ref{fig:cddf_main} are calculated from the posterior distribution directly.
We note that there are only two {\dla}s with {\mapval} $\lognhi > 22.5$ in our catalogue with high confidence ($\pdla > 0.99$).
The non-zero values in the {\cddf} are due to uncertainty in $\lognhi$, not positive detections.

\cite{Noterdaeme12} contains multi-{\dla}s, but, as described in Section 2.2 in their paper, they applied a stringent cut on their samples with {\cnr} $> 3$, where {\cnr} refers to the continuum-to-noise ratio.
The {\cddf} of {\nonetwo} in the Figure~\ref{fig:cddf_main} is thus a sub-sample of their catalogue.
We, on the other hand, use all data even those with low signal-to-noise ratios.
Comparing to our previously published {\cddf} \citep{Bird17},
the {\cddf} in this paper shows {\dla} detections at low $\nhi$ are consistent with \cite{Noterdaeme12}. Introducing the sub-{\dla} as an alternative model successfully regularises detections at $\sim 10^{20} \cm^{-2}$.\footnote{Note again the artifact at $\sim 10^{20} \cm^{-2}$ will not affect the analyses of $\dd N/\dd X$ or $\omegadla$ as the definition of a {\dla}s is absorbers with $\nhi > 10^{20.3} \cm^{-2}$.}

Figure~\ref{fig:dndx_main} shows the line density of {\dla}s.
Our results are again consistent with those of \cite{Prochaska2009} and \cite{Noterdaeme12} where they both agree.
Our detections are between those two catalogues at low redshift bins and consistent with \cite{Prochaska2009} in the highest redshift bin.
Comparing to our previous $\dd N/\dd X$ \citep{Bird17},
we moderately regularise the detections of {\dla}s at high redshifts.
This change shows that changing the mean model of the {\gp} to include the mean flux absorption prevents the pipeline confusing the suppression due to the Lyman alpha forest with a {\dla}. While the change of posterior modes in $\dd N/\dd X$ is large at high redshift bins,
we note that those changes are mostly within 95\% confidence interval of our previously published line densities.
All analyses shown measure a peak in $\dd N/ \dd X$ at $z \sim 3.5$. This may be partially due to $\zdla = 3.5$ the {\sdss} colour selection algorithm systematic identified by \cite{Prochaska:2009a}, which over-samples Lyman-limit systems (LLS), especially near the quasar, in the redshift range $3.0 - 3.6$ \citep{Worseck:2011, Fumagalli:2013}. Note however that in our analysis neighbouring redshift bins are highly correlated and so a statistical fluctuation is also a valid explanation. We have checked visually that our sub-DLA model successfully models spectra with a LLS in the proximate zone of the quasar emission peak.

Figure~\ref{fig:omega_dla_main} shows the total column density $\omegadla$ in {\dla}s in units of the cosmic density.
Our results are mostly consistent with \cite{Noterdaeme12} although we have slightly lower $\omegadla$ at $z\sim 2$. This is due to our Occam's razor penalty, which suppresses {\dla}s in spectra which are not long enough to include the full width of the {\dla}. Since these are all low redshift quasars, this suppresses {\dla} detections at $z < 2.3$.
As discussed in \cite{Noterdaeme12}, \cite{Ramirez2016}, and \cite{Bird17}, the relatively low $\omegadla$ of \cite{Prochaska2009} is due to the smaller sample size of the {\sdss} {\dr}5 dataset.
We also compare our $\omegadla$ to that measured by \cite{Crighton2015} at high redshifts ($z = 4$ and $z = 5$).
\cite{Crighton2015} used a small but higher signal-to-noise dataset.
Our results at $z = 4$ and $z = 5$ are consistent with those from \cite{Crighton2015}.
However, we note that the relatively small sample of \cite{Crighton2015} may bias it slightly low, as contributions from {\dla}s with $\nhi$ higher than expected to be in the survey will not be included in their $\omegadla$ estimate. Our Bayesian analysis includes possible contributions of undetected {\dla}s with column density up to $\lognhi = 23$ in the error bars via the prior on the column density.

Compared to our previously published $\omegadla$ \citep{Bird17}, we found a reduction in $\omegadla$ between $z = 4$ and $z = 5$.
This is due to the incorporation of a better mean flux vector model, which reduces the posterior density of high-column density systems for high-redshift absorbers (although within the $95\%$ confidence bars of the earlier work). Our confidence intervals are also substantially smaller for $\zdla \gtrsim 3.7$ than in \citep{Bird17}. This is due to our inclusion, for the first time, of information from the Lyman-$\beta$ absorption of the {\dla}s, which both constrains {\dla} properties and helps to distinguish {\dla}s from noise fluctuations.


We have tested the robustness of our method with respect to spectra with different {\snr}s and found that, as in \cite{Bird17}, the statistical properties predicted by our method are uncorrelated with the quasar {\snr}. Furthermore, the presence of a {\dla} is uncorrelated with the quasar redshift, fixing a statistical systematic in the earlier work.

As a cross-check of our wider catalogue, we also tested the {\cddf}, line densities, and total column densities of the {\dla}s in our catalogue with a full range of $\zdla$, from \Ly$\infty$ to \Ly$\alpha$. The {\cddf} was very similar to the {\cddf} excluding the Ly$\beta$ region shown in Figure~\ref{fig:cddf_main}, but with a moderate increase at high column density.
$\dd N/\dd X$ was almost identical to Figure~\ref{fig:dndx_main}, indicating that the detection of {\dla}s is robust even though we extend our sampling range to \Ly$\infty$. However, $\omegadla$ increases for $3.5 < \zdla < 4.0$. By visual inspection we found that this is due to the spectra where the quasar redshift from the {\sdss} pipeline in error and a Lyman break trough appears at the blue end of the spectrum in a region the code expects to contain only Ly$\beta$ absorption. As our model does not account for redshift errors, it explains the absorption due to these troughs by {\dla}s.

\subsection{Comparison to Garnett's Catalogue}

To understand the effect of the modifications we made to our model in this paper,
we visually inspected a subset of spectra with high model posteriors of a {\dla} in \cite{Garnett17} ($p_\textrm{DLA}^\textrm{Garnett}$) but low model posteriors in our current model ($p_\textrm{DLA}$). In particular, we chose spectra with $(p_\textrm{DLA}^\textrm{Garnett} - p_\textrm{DLA} > 0.99)$.

A large fraction of these spectra falls within the \Ly$\beta$ emission region. One plausible explanation is that the \Ly$\beta$ emission region has a higher noise variance, which makes it harder to distinguish the {\dla} and sub-{\dla} models.
We also checked that we are not unfairly preferring the sub-{\dla} model during model selection.
Our model selection uses the sub-{\dla} model only to regularise the {\dla} model and does not consider cases where {\dla}s and sub-{\dla}s occur in the same spectrum. Thus a spectrum with a clear detection of a sub-{\dla} could fail to detect a true {\dla} at a different redshift. In light of this,
we also tested if combining multi-{\dla} models with a sub-{\dla} affects our results.


We modified the {\dla} model, assuming that the {\dla} and sub-{\dla} models are independent, to include the sub-{\dla} model prior.
We then considered an iterative sampling procedure:
First, we sampled the $k^\textrm{th}$ {\dla} likelihood. Next we used the $k^\textrm{th}$ {\dla} parameter posterior as a prior
to sample $\mkdla$ and combine $\mkdla$ with the sub-{\dla} model via sampling a non-informative prior. The full procedure can be written as:
\begin{equation}
   \begin{split}
      p(\{\theta_i\}_{i=1}^k \mid \mkdla' &, \Data, \zqso)
      = \\
      (1 + p(&\theta_\textrm{sub} \mid \msubdla, \zqso))\times\\
      &p(\{\theta_i\}_{i=1}^k \mid \mkdla, \Data, \zqso),
   \end{split}
   \label{eq:parameter_prior_dla_subdlas}
\end{equation}
For computational simplicity, we only consider the modified model until $\mdla'{_{(3)}}$;
the probability of $\mdla'{_{(4)}}$ is expected to be insignificant comparing to the total {\dla} model posterior, $p(\{\mdla\} \mid \Data, \zqso)$.

In practice, however, we found that this made a small difference to our results, only marginally modifying the {\roc} curve and \cddf.
Moreover, the ability of the sub-{\dla} model to regularize low column density {\dla}s was reduced, so we have preserved our default model.

\subsection{Comparison to Parks Catalogue}
\begin{figure*}
   \includegraphics[width=2\columnwidth]{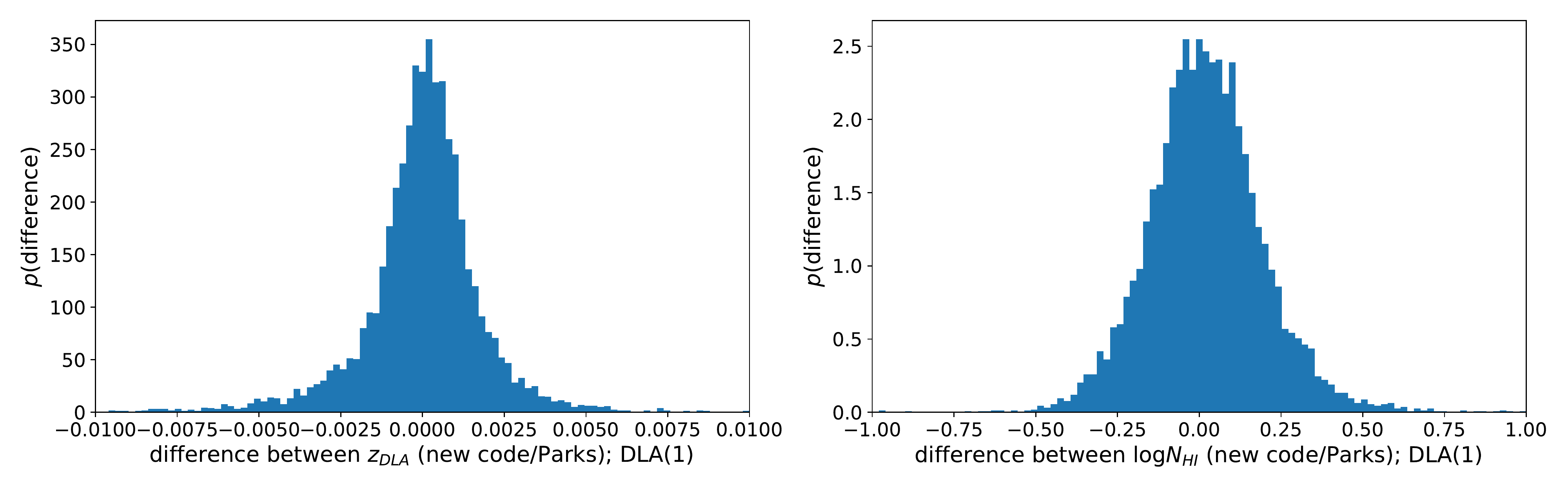}
   \caption{The difference between the {\mapval} estimates of the {\dla} parameters $\theta = (\zdla, \lognhi)$,
   against the predictions of \protect\cite{Parks18}.
   We consider spectra which both catalogues agree contain one {\dla}.}
   \label{fig:map_parks}
\end{figure*}

In this section, we compare our results with \cite{Parks18}.
We first show the differences between our {\mapval} predictions and Parks' predictions for {\dla} redshift and column density.
We required $p_\textrm{DLA}^\textrm{Parks} > 0.98$. We measured the difference in posterior parameters when both pipelines predicted one {\dla}.
As shown in Figure~\ref{fig:map_parks}, both histograms are roughly symmetric.
We measure small median offsets between two pipelines with
\begin{equation}
   \begin{split}
      \textrm{median}(z_\textrm{DLA}^\textrm{MAP} - z_\textrm{DLA}^\textrm{Parks}) &= 0.00010;\\
      \textrm{median}(\log{N_\textrm{HI}^\textrm{MAP}} - \log{N_\textrm{HI}^\textrm{Parks}}) &= 0.016.
   \end{split}
\end{equation}

We also compared our absorber redshift measurements and column density measurements to Parks' catalogue for those spectra which we both agree contain two {\dla}s.
The differences between these two have small median offsets of $\Delta \zdla = 0.000052$ and $\Delta \lognhi = 0.006$ (and dominated by low column density systems).

We show the disagreement between multi-{\dla} predictions for our catalogue and Parks' catalogue in Table~\ref{table:confusion_matrix}.
Note that though the multi-{\dla} detections between our method and Parks do not completely agree,
the level of disagreement is small: $6.1\%$. Moreover, if Parks predicts one or two {\dla}s, our method generally detects one or two {\dla}s.
There are however some spectra where we detected $> 2$ {\dla}s, but Parks detected none. To understand the statistical effect of this discrepancy, we compare our {\dla} properties to those reported by \cite{Parks18}. We plot the {\cddf} and $\dd N/ \dd X$ of that catalogue. We assume $p_\textrm{DLA}^\textrm{Parks} > 0.9$ represents a {\dla} and use $\zdla$ and $\lognhi$ reported in their catalogue in JSON format\footnote{\url{https://tinyurl.com/cnn-dlas}}. To compute the sightline path searched over, we assume their {\cnn} model was searching the range $\Ly \infty$ to $\Ly \alpha$ in the quasar rest-frame.
Note this differs slightly from \cite{Parks18} Section 3.2 where a sightline search radius ranging from $900${\AA} to $1346${\AA} in the quasar rest frame is given. However, we know the centers of {\dla}s should be at a redshift between $\Ly \infty$ and \Ly$\alpha$ in the rest frame and modify our search paths accordingly.


\begin{table*}
   \begin{tabular}{r | r r r r r}
      Parks & 0 DLA & 1 DLA & 2 DLAs & 3 DLAs & 4 DLAs \\
      Garnett with Multi-DLAs & & & & & \\
      \hline\hline
      0 DLA  & {\bf 138726} & 6197  & 142    & 6      & 0      \\
      1 DLA  & 3050   & {\bf 8752}  & 335    & 4      & 0      \\
      2 DLAs & 293    & 570   & {\bf 566}    & 28     & 0      \\
      3 DLAs & 30     & 39    & 34     & {\bf 21}     & 0      \\
      4 DLAs & 5      & 9     & 6      & 1            & 0
   \end{tabular}
   \caption{The confusion matrix for multi-{\dla}s detections between Garnett with multi-{\dla}s and Parks.
   Note we require both the model posteriors in Garnett and {\dla} confidence in Parks to be larger than $0.98$.
   We also require $\lognhi > 20.3$.}
   \label{table:confusion_matrix}
\end{table*}

\begin{figure}
   \centering
   \includegraphics[width=1\columnwidth]{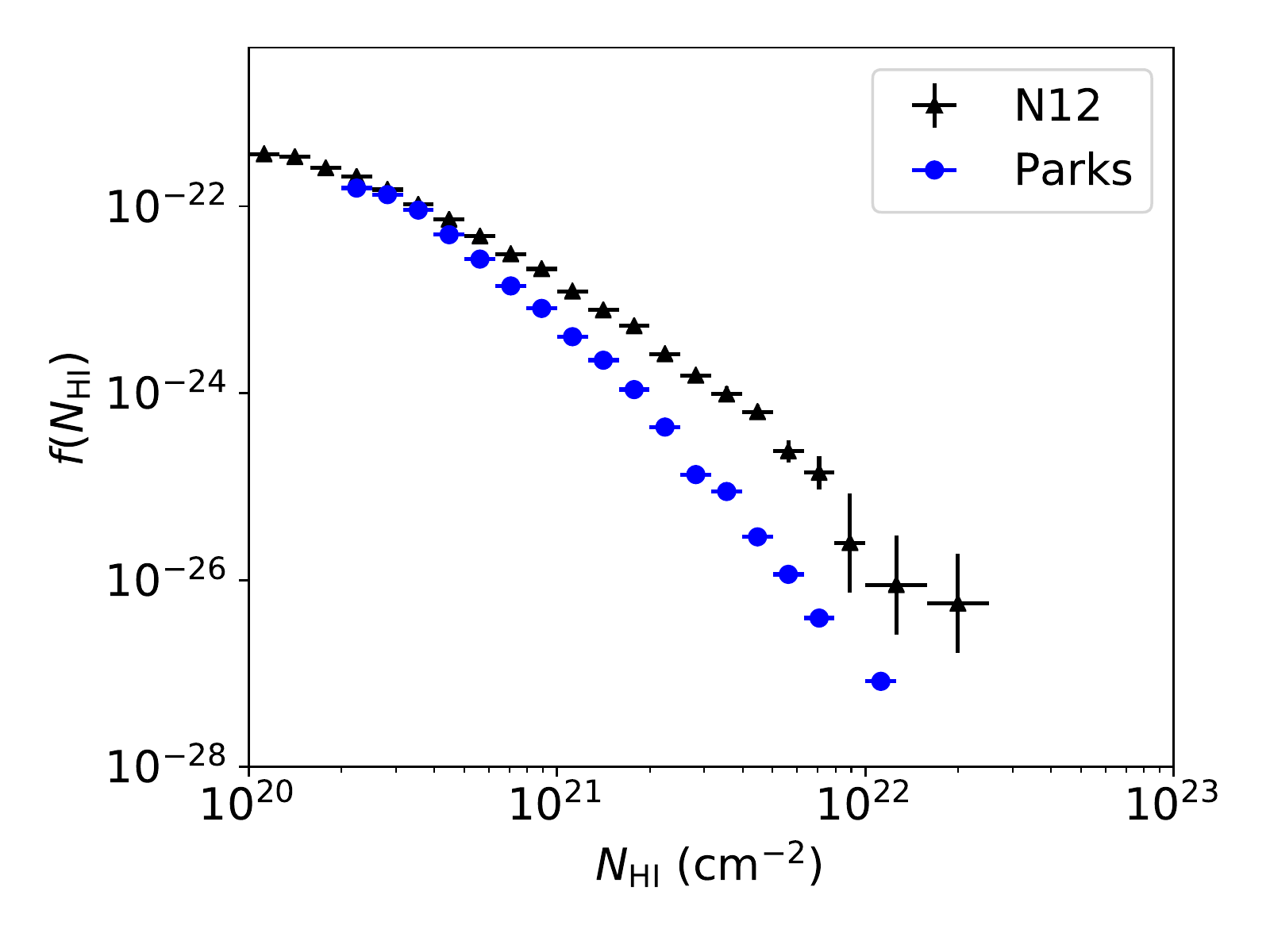}
   \caption{The column density distribution function from \protect\cite{Parks18}, showing that the {\cnn} algorithm substantially underestimates the number of {\dla}s in the high-$\nhi$ regime.}
   \label{fig:cddf_parks}
\end{figure}

\begin{figure}
   \centering
   \includegraphics[width=1\columnwidth]{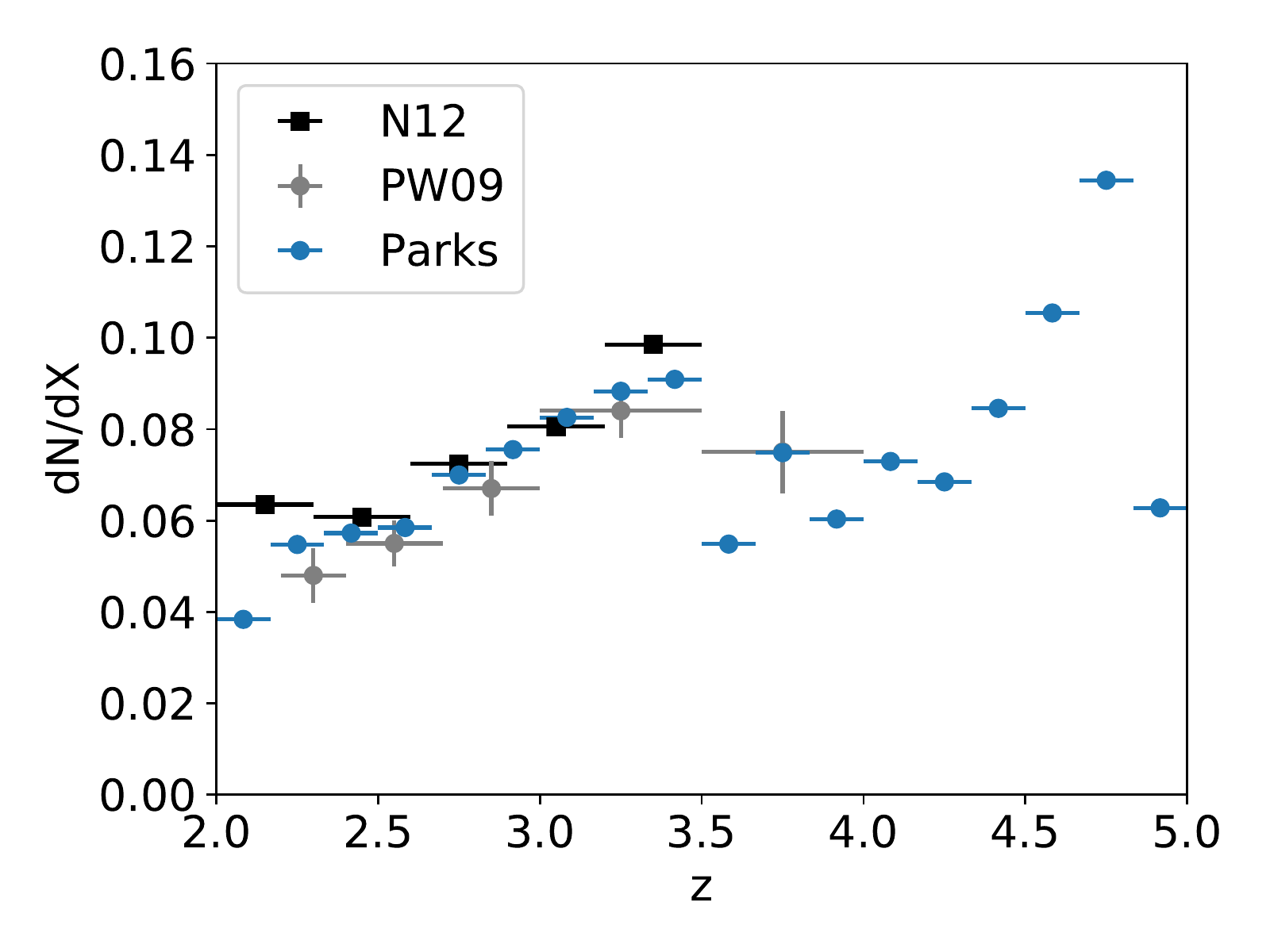}
   \caption{$\dd N/\dd X$ from \protect\cite{Parks18}. The $\dd N/\dd X$ agrees well with other surveys, but there is a moderate deficit of {\dla}s at high redshifts.}
   \label{fig:dndx_parks}
\end{figure}

Figure~\ref{fig:dndx_parks} shows that $\dd N/ \dd X$ is consistent with \cite{Noterdaeme12} for $\zdla < 3.5$ (although lower than our measurement at higher redshift). The {\cnn} is thus successfully detecting {\dla}s, especially the most common case of {\dla}s with a low column density. There are fewer {\dla}s detected at higher redshift, likely reflecting the increased difficulty for the {\cnn} of distinguishing {\dla}s from the \Lya~forest. This is discussed in \cite{Parks18}, who note that the \cnn~finds it difficult to detect a weak {\dla} in noisy spectra.
However, as shown in Figure~\ref{fig:cddf_parks}, the \cddf~measured by the {\cnn} model is significantly discrepant with other surveys for large column densities. Note that the scale is logarithmic: the \cnn~is failing to detect $> 60\%$ of {\dla}s with $\lognhi > 21$. We noticed that large {\dla}s were often split into two objects with lower column density, which accounts for many of the discrepancies between our two datasets. We suspect this might be due to the limited size of the convolutional filters used by \cite{Parks18}. If the filter is not large enough to contain the full damping wings of a given {\dla}, the allowed column density would be artificially limited.


\section{Conclusion}
We have presented a revised pipeline for detecting {\dla}s in {\sdss} quasar spectra based on \cite{Garnett17}. We have extended the pipeline to reliably detect up to $4$ {\dla}s per spectrum. We have performed modifications to our model for the \Lya~forest to improve the reliability of {\dla} detections at high redshift and introduced a model for sub-{\dla}s to improve our measurement of low column density {\dla}s. Finally we introduced a penalty on the {\dla} model based on Occam's razor which meant that spectra for which both models are a poor fit generally prefer the no-{\dla} model.

Our results include a public {\dla} catalogue, with several examples shown above and further examples easily plotted using a python package. We have visually inspected several extreme cases to validate our results and compared extensively to several earlier {\dla} catalogues: the {\dr}9 concordance catalogue \citep{Lee2013} and a {\dr}12 catalogue using a {\cnn} \citep{Parks18}. Our new pipeline had very good performance validated against both catalogues.

Based on the revised pipeline, we also presented a new measurement of the abundance of neutral hydrogen from $z = 2$ to $z = 5$ using similar calculations to \cite{Bird17}. The statistical properties of {\dla}s were in good agreement with our previous results \citep{Bird17} and consistent with \cite{Noterdaeme12}, \cite{Prochaska2009}, and \cite{Crighton2015}. The modifications made, including introducing a sub-{\dla} model, adjusting the mean flux, and penalizing complex models with Occam's razor, remove over-detections of low column density absorbers and make more robust predictions for the properties of {\dla}s at $z > 4$. Similarly to previous work, we detect only a small increase in the {\cddf} for $2 < z < 4$, and a similarly moderate increase in the line density of {\dla}s and $\omegadla$ over this redshift range.

\section*{Acknowledgements}

The authors thank Bryan Scott and Reza Monadi for valuable comments.
This work was partially supported by an Amazon Machine Learning Research allocation on EC2 and UCR's HPCC.
SB was supported by the National Science Foundation (NSF) under award number AST-1817256.
RG was supported by the NSF under award number IIS–1845434.

\section*{Data availability}
All the code to reproduce the data products is available in our GitHub repo: \url{https://github.com/rmgarnett/gp_dla_detection/tree/master/multi_dlas}.
The final data products are available in this Google Drive: \url{http://tiny.cc/multidla_catalog_gp_dr12q},
including a MAT (HDF5) catalogue and a JSON catalogue.
README files are included in each folder to explain the content of the catalogues.

\bibliography{sample}

\appendix

\section[Sample posteriors for MDLA2]{Sample posteriors for $\mdla{_{(2)}}$}

The calculation of the Poisson-Binomial process in Eq.~\ref{eq:poission_binomial} computes the probability of N {\dla}s within a given column density or redshift bin on the sample posteriors $\pidla(\theta) = p(\{\mdla\} \mid \yvec_i, \lambdavec_i, \nuvec_i, \theta, \zqso{_i})$, where $i$ represents the index of the quasar sample.
However, with more than 1 {\dla}, we will not just have parameters in two-dimensions $\theta = (\lognhi, \zdla)$ but also parameters from the second or third {\dla}s $\{\theta_j\}_{j=1}^k = \{(\lognhi{_j}, \zdla{_j})\}_{j=1}^k$, with $k$ {\dla}s.
It is thus not straightforward to see how we can calculate the Poisson-Binomial process on the sample posteriors with more than 1 {\dla}.

Here we provide a procedure to calculate the sample posteriors of the second {\dla} given the parameters of the first {\dla}.
The sample posteriors of the second {\dla} could be written as:
\begin{equation}
   \begin{split}
      p(&\textrm{2nd DLA at } \theta = (\lognhi, \zdla))\\
      &=
      \int_{\textrm{1st DLA} \in \theta'}
      p(\textrm{1st DLA at }\theta'\textrm{ and 2nd DLA at }\theta)\dd\theta'\\
      &=
      \int_{\theta'} p(\theta, \theta' \mid \mdla{_{(2)}}, \Data) \dd \theta'
   \end{split}
\end{equation}
where we marginalize the first {\dla} at parameters $\theta' = (\lognhi', \zdla')$ with a given 2nd {\dla} parameter $\theta = (\lognhi, \zdla)$.
We can furthermore write the joint posterior density into a likelihood density using Bayes rule:
\begin{equation}
   \begin{split}
      p(\theta, &\theta' \mid \mdla{_{(2)}}, \Data)
      \\
      \propto &p(\Data \mid \theta, \theta', \mdla{_{(2)}})
      p(\theta', \theta \mid \mdla{_{(2)}})\\
      = &p(\Data \mid \theta, \theta', \mdla{_{(2)}})
      p(\theta \mid \mdla{_{(1)}}) \\
      &p(\theta' \mid \mdla{_{(1)}}, \Data)
   \end{split}
\end{equation}
where the joint prior density $p(\theta', \theta \mid \mdla{_{(2)}})$ could be written as a product of a non-informative prior and an informed prior.

The posterior density of the second {\dla} could thus be expressed as a discrete sum over $\theta'$ at the informed prior density:
\begin{equation}
   \begin{split}
      p(\textrm{2nd } &\textrm{DLA at } \theta)\\
      \propto \int_{\theta'}
      &p(\Data \mid \theta, \theta', \mdla{_{(2)}})
      p(\theta \mid \mdla{_{(1)}}) \\
      &p(\theta' \mid \mdla{_{(1)}}, \Data)\dd \theta'\\
      \simeq
      \frac{1}{N}
      &\sum_{i=1}^N
      p(\Data \mid \theta, \theta'_i, \mdla{_{(2)}})
      p(\theta \mid \mdla{_{(1)}}),
   \end{split}
\end{equation}
where
\begin{equation}
   \theta'_i \sim p(\theta' \mid \mdla{_{(1)}}, \Data).
\end{equation}

However, for each $\theta$, we only have one $\theta'$.
We thus can simplify the discrete sum as:
\begin{equation}
   \begin{split}
      p(&\textrm{2nd DLA at } \theta)\\
      &\propto p(\Data \mid \theta, \theta'_i, \mdla{_{(2)}})
      p(\theta \mid \mdla{_{(1)}}),
   \end{split}
\end{equation}
where the non-informative prior $p(\theta \mid \mdla{_{(1)}})$ expresses the way we sample $\theta$ for $p(\textrm{2nd DLA at } \theta)$.

To get the normalized posterior density for the 2nd {\dla}, we can directly normalize on the joint likelihood density:
\begin{equation}
   \begin{split}
      p(&\textrm{2nd DLA at } \theta)\\
      &= \frac{p(\Data \mid \{\theta, \theta'\}_j, \mdla{_{(2)}})}{
         \sum_{j=1}^N p(\Data \mid \{\theta, \theta'\}_j, \mdla{_{(2)}})
      }\\
      &=\frac{p(\Data \mid \{\theta, \theta'\}_j, \mdla{_{(2)}})}{
         N^2 \frac{1}{N^2}\sum_{j=1}^N p(\Data \mid \{\theta, \theta'\}_j, \mdla{_{(2)}})
      }\\
      &= \frac{1}{N^2}
      \frac{p(\Data \mid \{\theta, \theta'\}_j, \mdla{_{(2)}})}{
         p(\Data \mid \mdla{_{(2)}})
      }
   \end{split}
\end{equation}

We thus can compute the posterior density for the first {\dla} and second {\dla} at a given $\theta$:
\begin{equation}
   \begin{split}
      p(\textrm{1 or } &\textrm{2 DLAs at }\theta)\\
      = &\Prob(\mdla{_{(1)}}) p(\textrm{1st DLA at } \theta)\\
      &+ \Prob(\mdla{_{(2)}}) p(\textrm{2nd DLA at } \theta)
   \end{split}
\end{equation}

%
%

\section[Tables for CDDF, dN/dX and OmegaDLA]{Tables for CDDF, dN/dX, and $\omegadla$}
\begin{table*}
   \centering
  \begin{tabular}{cccccc}
   \hline
   $\log_{10} \mathrm{N}_\mathrm{HI}$ & $f(N_\mathrm{HI})$  $( 10^{ -21 } )$ & $68$\% limits $( 10^{ -21 } )$ & $95$\% limits $( 10^{ -21 } )$ \\
    \hline
   $ 20.0 - 20.1 $ & $ 0.413 $ & $ 0.405 - 0.422 $ & $ 0.398 - 0.430 $  \\
   $ 20.1 - 20.2 $ & $ 0.241 $ & $ 0.235 - 0.247 $ & $ 0.229 - 0.252 $  \\
   $ 20.2 - 20.3 $ & $ 0.175 $ & $ 0.171 - 0.180 $ & $ 0.167 - 0.184 $  \\
   $ 20.3 - 20.4 $ & $ 0.136 $ & $ 0.132 - 0.139 $ & $ 0.129 - 0.142 $  \\
   $ 20.4 - 20.5 $ & $ 0.101 $ & $ [9.88  - 10.41 ]\times 10^{ -2 }$ & $ [9.63  - 10.67 ]\times 10^{ -2 }$  \\
   $ 20.5 - 20.6 $ & $ 7.60 \times 10^{ -2 }$ & $ [7.40  - 7.80 ]\times 10^{ -2 }$ & $ [7.21  - 8.00 ]\times 10^{ -2 }$  \\
   $ 20.6 - 20.7 $ & $ 5.20 \times 10^{ -2 }$ & $ [5.06  - 5.36 ]\times 10^{ -2 }$ & $ [4.91  - 5.50 ]\times 10^{ -2 }$  \\
   $ 20.7 - 20.8 $ & $ 3.84 \times 10^{ -2 }$ & $ [3.73  - 3.96 ]\times 10^{ -2 }$ & $ [3.63  - 4.07 ]\times 10^{ -2 }$  \\
   $ 20.8 - 20.9 $ & $ 2.52 \times 10^{ -2 }$ & $ [2.44  - 2.60 ]\times 10^{ -2 }$ & $ [2.37  - 2.69 ]\times 10^{ -2 }$  \\
   $ 20.9 - 21.0 $ & $ 1.67 \times 10^{ -2 }$ & $ [1.61  - 1.72 ]\times 10^{ -2 }$ & $ [1.56  - 1.78 ]\times 10^{ -2 }$  \\
   $ 21.0 - 21.1 $ & $ 1.03 \times 10^{ -2 }$ & $ [9.94  - 10.75 ]\times 10^{ -3 }$ & $ [9.57  - 11.15 ]\times 10^{ -3 }$  \\
   $ 21.1 - 21.2 $ & $ 7.21 \times 10^{ -3 }$ & $ [6.92  - 7.49 ]\times 10^{ -3 }$ & $ [6.65  - 7.78 ]\times 10^{ -3 }$  \\
   $ 21.2 - 21.3 $ & $ 4.17 \times 10^{ -3 }$ & $ [3.99  - 4.37 ]\times 10^{ -3 }$ & $ [3.81  - 4.56 ]\times 10^{ -3 }$  \\
   $ 21.3 - 21.4 $ & $ 2.87 \times 10^{ -3 }$ & $ [2.74  - 3.01 ]\times 10^{ -3 }$ & $ [2.62  - 3.14 ]\times 10^{ -3 }$  \\
   $ 21.4 - 21.5 $ & $ 1.49 \times 10^{ -3 }$ & $ [1.41  - 1.58 ]\times 10^{ -3 }$ & $ [1.33  - 1.68 ]\times 10^{ -3 }$  \\
   $ 21.5 - 21.6 $ & $ 8.71 \times 10^{ -4 }$ & $ [8.23  - 9.31 ]\times 10^{ -4 }$ & $ [7.76  - 9.79 ]\times 10^{ -4 }$  \\
   $ 21.6 - 21.7 $ & $ 4.03 \times 10^{ -4 }$ & $ [3.74  - 4.41 ]\times 10^{ -4 }$ & $ [3.46  - 4.69 ]\times 10^{ -4 }$  \\
   $ 21.7 - 21.8 $ & $ 2.15 \times 10^{ -4 }$ & $ [1.96  - 2.37 ]\times 10^{ -4 }$ & $ [1.77  - 2.60 ]\times 10^{ -4 }$  \\
   $ 21.8 - 21.9 $ & $ 1.41 \times 10^{ -4 }$ & $ [1.29  - 1.56 ]\times 10^{ -4 }$ & $ [1.17  - 1.70 ]\times 10^{ -4 }$  \\
   $ 21.9 - 22.0 $ & $ 4.75 \times 10^{ -5 }$ & $ [4.04  - 5.70 ]\times 10^{ -5 }$ & $ [3.56  - 6.41 ]\times 10^{ -5 }$  \\
   $ 22.0 - 22.1 $ & $ 2.08 \times 10^{ -5 }$ & $ [1.70  - 2.64 ]\times 10^{ -5 }$ & $ [1.32  - 3.21 ]\times 10^{ -5 }$  \\
   $ 22.1 - 22.2 $ & $ 8.99 \times 10^{ -6 }$ & $ [7.49  - 13.49 ]\times 10^{ -6 }$ & $ [6.00  - 16.49 ]\times 10^{ -6 }$  \\
   $ 22.2 - 22.3 $ & $ 4.76 \times 10^{ -6 }$ & $ [3.57  - 7.14 ]\times 10^{ -6 }$ & $ [2.38  - 9.52 ]\times 10^{ -6 }$  \\
   $ 22.3 - 22.4 $ & $ 1.89 \times 10^{ -6 }$ & $ [9.46  - 37.83 ]\times 10^{ -7 }$ & $ [9.46  - 56.74 ]\times 10^{ -7 }$  \\
   $ 22.4 - 22.5 $ & $ 3.00 \times 10^{ -6 }$ & $ [2.25  - 3.76 ]\times 10^{ -6 }$ & $ [1.50  - 4.51 ]\times 10^{ -6 }$  \\
   $ 22.5 - 22.6 $ & $ 5.97 \times 10^{ -7 }$ & $ [5.97  - 17.90 ]\times 10^{ -7 }$ & $ [5.97  - 29.83 ]\times 10^{ -7 }$  \\
   $ 22.6 - 22.7 $ & $0$ & $0 -  9.48 \times 10^{ -7 }$ & $0 -  9.48 \times 10^{ -7 }$  \\
   $ 22.7 - 22.8 $ & $ 3.76 \times 10^{ -7 }$ & $ [3.76  - 11.29 ]\times 10^{ -7 }$ & $ [3.76  - 11.29 ]\times 10^{ -7 }$  \\
   $ 22.8 - 22.9 $ & $0$ & $0 -  2.99 \times 10^{ -7 }$ & $0 -  5.98 \times 10^{ -7 }$  \\
   $ 22.9 - 23.0 $ & $0$ & $0 -  2.38 \times 10^{ -7 }$ & $0 -  4.75 \times 10^{ -7 }$  \\
   \hline
   \end{tabular}
   \caption{Average column density distribution function for all {\dla}s with $2 < z < 5$.
   The table is generated by using $p(\{\mdla\} \mid \yvec, \lambdavec, \nuvec, \zqso)$.
   See also Figure~\ref{fig:cddf_main}.}
  \label{tab:cddf_all.txt}
\end{table*}

\begin{table*}
   \centering
  \begin{tabular}{cccccc}
   \hline
   $z$ & dN/dX & $68$\% limits & $95$\% limits \\
    \hline
   $ 2.00 - 2.17 $ & $ 0.0309 $ & $ 0.0302 - 0.0317 $ & $ 0.0294 - 0.0325 $  \\
   $ 2.17 - 2.33 $ & $ 0.0448 $ & $ 0.0438 - 0.0458 $ & $ 0.0428 - 0.0468 $  \\
   $ 2.33 - 2.50 $ & $ 0.0497 $ & $ 0.0485 - 0.0510 $ & $ 0.0474 - 0.0521 $  \\
   $ 2.50 - 2.67 $ & $ 0.0528 $ & $ 0.0514 - 0.0542 $ & $ 0.0501 - 0.0556 $  \\
   $ 2.67 - 2.83 $ & $ 0.0676 $ & $ 0.0658 - 0.0694 $ & $ 0.0641 - 0.0711 $  \\
   $ 2.83 - 3.00 $ & $ 0.0721 $ & $ 0.0700 - 0.0743 $ & $ 0.0680 - 0.0764 $  \\
   $ 3.00 - 3.17 $ & $ 0.0760 $ & $ 0.0734 - 0.0788 $ & $ 0.0709 - 0.0814 $  \\
   $ 3.17 - 3.33 $ & $ 0.0846 $ & $ 0.0811 - 0.0885 $ & $ 0.0779 - 0.0919 $  \\
   $ 3.33 - 3.50 $ & $ 0.0824 $ & $ 0.0785 - 0.0868 $ & $ 0.0747 - 0.0910 $  \\
   $ 3.50 - 3.67 $ & $ 0.0835 $ & $ 0.0786 - 0.0888 $ & $ 0.0737 - 0.0937 $  \\
   $ 3.67 - 3.83 $ & $ 0.0738 $ & $ 0.0671 - 0.0806 $ & $ 0.0618 - 0.0873 $  \\
   $ 3.83 - 4.00 $ & $ 0.0594 $ & $ 0.0512 - 0.0675 $ & $ 0.0454 - 0.0757 $  \\
   $ 4.00 - 4.17 $ & $ 0.0665 $ & $ 0.0570 - 0.0797 $ & $ 0.0475 - 0.0892 $  \\
   $ 4.17 - 4.33 $ & $ 0.1033 $ & $ 0.0893 - 0.1228 $ & $ 0.0726 - 0.1396 $  \\
   $ 4.33 - 4.50 $ & $ 0.0966 $ & $ 0.0805 - 0.1208 $ & $ 0.0644 - 0.1409 $  \\
   $ 4.50 - 4.67 $ & $ 0.1137 $ & $ 0.0885 - 0.1453 $ & $ 0.0632 - 0.1706 $  \\
   $ 4.67 - 4.83 $ & $ 0.1131 $ & $ 0.0754 - 0.1634 $ & $ 0.0503 - 0.2011 $  \\
   $ 4.83 - 5.00 $ & $0$ & $0 -  0.057 $ & $0 -  0.085 $  \\
   \hline
   \end{tabular}
   \caption{Table of $\dd N/\dd X$ values from our multi-{\dla} catalogue for $2 < z < 5$.
   The table is generated by using $p(\{\mdla\} \mid \yvec, \lambdavec, \nuvec, \zqso)$.
   See also Figure~\ref{fig:dndx_main}.}
  \label{tab:dndx_all.txt}
\end{table*}

\begin{table*}
   \centering
  \begin{tabular}{cccccc}
   \hline
   $z$ & $\Omega_\mathrm{DLA} (10^{-3}) $ & $68$\% limits & $95$\% limits \\
    \hline
   $ 2.00 - 2.17 $ & $ 0.385 $ & $ 0.371 - 0.400 $ & $ 0.358 - 0.416 $  \\
   $ 2.17 - 2.33 $ & $ 0.532 $ & $ 0.516 - 0.550 $ & $ 0.501 - 0.568 $  \\
   $ 2.33 - 2.50 $ & $ 0.645 $ & $ 0.620 - 0.679 $ & $ 0.596 - 0.720 $  \\
   $ 2.50 - 2.67 $ & $ 0.653 $ & $ 0.624 - 0.689 $ & $ 0.598 - 0.728 $  \\
   $ 2.67 - 2.83 $ & $ 0.786 $ & $ 0.759 - 0.814 $ & $ 0.732 - 0.841 $  \\
   $ 2.83 - 3.00 $ & $ 0.792 $ & $ 0.764 - 0.822 $ & $ 0.737 - 0.850 $  \\
   $ 3.00 - 3.17 $ & $ 0.910 $ & $ 0.865 - 0.972 $ & $ 0.826 - 1.046 $  \\
   $ 3.17 - 3.33 $ & $ 1.051 $ & $ 1.002 - 1.101 $ & $ 0.957 - 1.154 $  \\
   $ 3.33 - 3.50 $ & $ 0.958 $ & $ 0.891 - 1.031 $ & $ 0.829 - 1.106 $  \\
   $ 3.50 - 3.67 $ & $ 1.297 $ & $ 1.220 - 1.380 $ & $ 1.147 - 1.455 $  \\
   $ 3.67 - 3.83 $ & $ 1.303 $ & $ 1.222 - 1.391 $ & $ 1.144 - 1.486 $  \\
   $ 3.83 - 4.00 $ & $ 0.891 $ & $ 0.742 - 1.052 $ & $ 0.639 - 1.205 $  \\
   $ 4.00 - 4.17 $ & $ 0.993 $ & $ 0.746 - 1.245 $ & $ 0.564 - 1.488 $  \\
   $ 4.17 - 4.33 $ & $ 1.519 $ & $ 1.341 - 1.718 $ & $ 1.168 - 1.923 $  \\
   $ 4.33 - 4.50 $ & $ 1.085 $ & $ 0.880 - 1.325 $ & $ 0.702 - 1.695 $  \\
   $ 4.50 - 4.67 $ & $ 1.741 $ & $ 1.282 - 2.224 $ & $ 0.666 - 2.851 $  \\
   $ 4.67 - 4.83 $ & $ 1.239 $ & $ 0.826 - 1.712 $ & $ 0.484 - 2.222 $  \\
   $ 4.83 - 5.00 $ & $0$ & $0 -  0.213 $ & $0 -  0.492 $  \\
   \hline
    \end{tabular}
   \caption{Table of $\Omega_\mathrm{DLA}$ values.
   The table is generated by using $p(\{\mdla\} \mid \yvec, \lambdavec, \nuvec, \zqso)$.
   See also Figure~\ref{fig:omega_dla_main}.}
  \label{tab:omega_dla_all.txt}
\end{table*}

\label{lastpage}

\end{document}